\documentclass{aa}  

\usepackage{graphicx}
\usepackage[varg]{txfonts}
\bibpunct{(}{)}{;}{a}{}{,}

\begin{document}

   \title{Revealing the pulsational properties of the V777 Her star \mbox{KUV 05134+2605} by its long-term monitoring
   }
   
   \authorrunning{Zs. Bogn\'ar et al.}

   \titlerunning{Long-term monitoring of \mbox{KUV 05134+2605}}

   \author{Zs. Bogn\'ar
          \inst{1}
          \and
           M. Papar\'o\inst{1}
          \and
           A. H. C\'orsico\inst{2,3} 
          \and
	  S. O. Kepler\inst{4}
	  \and
	  \'A. Gy\H orffy\inst{5}
          }

   \institute{
   Konkoly Observatory, MTA CSFK, Konkoly Thege M. u. 15-17, H--1121 Budapest, Hungary\\
   \email{bognar@konkoly.hu}
   \and
   Facultad de Ciencias Astron\'omicas y Geof\'{\i}sicas, Universidad Nacional de La Plata, Argentina
   \and
   Consejo Nacional de Investigaciones Cient\'{\i}ficas y T\'enicas (CONICET), Argentina
   \and
   Instituto de F\'{\i}sica, Universidade Federal do Rio Grande do Sul, 91501-900 Porto-Alegre, RS, Brazil
   \and
   Department of Astronomy, E\"otv\"os Lor\'and University, P\'azm\'any P\'eter s\'et\'any 1/A, H-1117 Budapest, Hungary
              }

   \date{Received  ; accepted }

 
  \abstract
   {\mbox{\object{KUV 05134+2605}} is one of the 21 pulsating DB white dwarfs (V777 Her or DBV variables) 
   known so far. The detailed investigation of the short-period and low-amplitude pulsations of these relatively 
   faint targets requires considerable observational efforts from the ground, long-term single-site or multisite 
   observations. The observed amplitudes of excited modes undergo short-term variations in many cases, which makes 
   the determination of pulsation modes difficult.}
   {We aim to determine the pulsation frequencies of \mbox{KUV 05134+2605}, find regularities between the 
   frequency and period components, and perform its asteroseismic investigation for the first time.}
   {We re-analysed the data already published, and collected new measurements. We compared the frequency 
   content of the different datasets from the different epochs and performed various tests to check the 
   reliability of the frequency determinations. The mean period spacings were investigated with 
   linear fits to the observed periods, Kolmogorov-Smirnov and Inverse Variance significance tests, and Fourier 
   analysis of different period sets, including a Monte Carlo test simulating the effect of alias ambiguities. We 
   employed fully evolutionary DB white dwarf models for the asteroseismic investigations.}
   {We identified 22 frequencies between $1280$ and $2530\,\mu$Hz. These form 12 groups, 
   which suggests at least 12 possible frequencies for the asteroseismic investigations. Thanks to the extended 
   observations, \mbox{KUV 05134+2605} joined the group of rich white dwarf pulsators. 
   We identified one triplet and at least one doublet with a $\approx9\,\mu$Hz frequency separation, from which
   we derived a stellar rotation period of $0.6$\,d. We determined the mean period spacings of $\approx 31$ and $18$\,s 
   for the modes we propose as dipole and quadrupole, respectively.
   We found an excellent agreement between the stellar mass derived from the $\ell=1$ period spacing and the 
   period-to-period fits, all providing $M_*= 0.84-0.85\,M_{\odot}$ solutions. Our study suggests that \mbox{KUV 05134+2605} 
   is the most massive amongst the known V777 Her stars.}
   {}


   \maketitle
%

\section{Introduction}

The discovery of the first pulsating DB-type white dwarf (\object{GD\,358} or V777 Her) was not serendipitous: 
it was the result of a dedicated search for unstable modes in these stars, driven by the partial ionization of 
the atmospheric helium \citep{1982ApJ...262L..11W}.

The V777 Her stars are similar to their `cousins', the ZZ Ceti (or DAV) stars, in their pulsational properties. Both show light 
variations due to non-radial $g$-mode pulsations with periods between $100-1400$\,s. However, the V777 Her stars have larger 
effective temperatures ($22\,000 - 29\,000$\,K) than the DAVs ($10\,900 - 12\,300$\,K). 
Another significant difference is the 
composition of their atmospheres: hydrogen dominates the DAV, while helium constitute the DBV atmospheres. 
We note that the classification of the recently discovered cool ($T_\mathrm{eff}\sim8000-9000$\,K), extremely low-mass 
pulsating DA white dwarfs
as ZZ Ceti stars was proposed by the modelling results of \citet{2013ApJ...762...57V}. However, except their similar
driving mechanism and surface composition, they are completely different, e.g. they likely have binary origin, their
cores are dominated by helium, and they pulsate with longer periods associated to $g$-modes than the ZZ Ceti stars 
\citep{2013MNRAS.436.3573H}.
About $80\%$ of 
the white dwarfs belong to the DA spectral class (e.g. \citealt{2013ApJS..204....5K}) and most of the known 
white dwarf pulsators are ZZ Ceti variables. With the discovery of pulsations in \object{KIC\,8626021}, the number of known 
V777 Her stars has reached only 21 \citep{2011ApJ...736L..39O}.

The regularities between the observed frequencies and periods offer great potential for the asteroseismic investigations 
of pulsating white dwarfs. Theoretically, rotational splitting of dipole and quadrupole modes into three or five 
equally spaced components allow the determination of the stellar rotation period, but the internal differential rotation 
\citep{1999ApJ...516..349K} and the presence of the magnetic field \citep{1989ApJ...336..403J} can complicate this picture. 
Finding these rotationally split frequencies also can be complicated: we need a long enough dataset to resolve the components, 
the overlapping splitting structures of the closely spaced modes can make the finding of the regularities difficult, and the 
frequency separations can be close to the daily alias of ground-based measurements. Furthermore, observational results indicate 
that in many cases not all of the components reach an observable amplitude at a given time, for unknown reasons 
\citep{2008ARA&A..46..157W}. 

The other theoretically predicted regularity is the quasi equally spaced formation of the consecutive radial overtone 
pulsation periods having the same spherical harmonic index ($\ell$). The inner chemical stratification 
breaks the uniform spacings, while the mean period spacing is an indicator of the total stellar mass.

Typical feature of the white dwarf pulsation is that the light curves are not sinusoidal, and we can detect combination
($nf_i \pm mf_j$), harmonic ($nf_i$), or even subharmonic ($n/2f_i$) peaks in the corresponding Fourier transforms. 
These features reveal the importance of the non-linear effects on the observed light variations.

For further details on the general properties of white dwarf pulsators, we refer to the reviews of \citet{2008ARA&A..46..157W}, 
\citet{2008PASP..120.1043F} and \citet{2010A&ARv..18..471A}.

The pulsation frequencies of V777 Her stars are reported to undergo short-term 
amplitude and phase variations (see e.g. Handler et al. 2013 and \citealt{2002MNRAS.335..698H}).
The most famed example is the so-called sforzando effect 
observed in GD\,358. This star showed a high-amplitude sinusoidal light variation for a short period of time, instead of the 
non-linear variability observed just a day before (\citealt{2003A&A...401..639K}, \citealt{2009ApJ...693..564P}). To find an 
explanation for the short-term changes in the pulsational behaviour of a star is challenging. As convection plays an important 
role in the driving of pulsations, the pulsation/convection interactions could explain some of the temporal variations 
\citep{2010ApJ...716...84M}. In the case of the pre-white dwarf GW Vir pulsator \object{PG 0122+200}, \citet{2011A&A...528A...5V} 
found that the temporal variations of pulsation modes may be the result of resonant coupling induced by rotation within triplets. 
In some cases, observed amplitude variations are the result of the insufficient frequency resolution of the datasets, and do not 
reflect real changes in the frequencies' energy content.

We assume that the structure of a star, which determines which pulsation frequencies can be excited, does not change 
on such short time scales as the number of detectable frequencies and their amplitudes in many white dwarfs. 
As a consequence of amplitude variations, one dataset reveals only a subset of the possible frequencies, but
with observations at different epochs, we can add more and more frequencies to the list of observed ones, essential 
for asteroseismic investigations (see e.g. the pioneer case of \object{G29-38} by \citealt{1998ApJ...495..424K}).
Here, observations of \mbox{KUV 05134+2605} at different epochs provides a more complete sampling of the 
pulsation frequencies, which will then lead to a more detailed asteroseismic analysis.

Asteroseismology can provide information on the masses of the envelope layers, the core structure and the stellar mass. These 
are crucial for building more realistic models and improving our knowledge of stellar evolution. Since more than $95\%$ 
of the stars will end their life as white dwarfs, and the information we gather on pulsating members is representative of the 
non-pulsating ones -- as most, if not all stars pulsate when they cross the instability strip --, asteroseismic studies of 
DAVs and DBVs provide important constraints on our models of stellar evolution and galactic history.

In the case of \mbox{KUV 05134+2605}, previous observations suggested a rich amplitude spectrum showing considerable amplitude 
variations from season to season \citep{2003MNRAS.340.1031H}. This made the star a promising target for further monitoring. Our 
main goal was to find the excited normal modes with data obtained in different seasons. In this paper, we discuss our findings 
and the results on the asteroseismic investigations based on the set of pulsation modes. 

\section[]{Observations and data reduction}

The first detection of the light variations of \mbox{KUV 05134+2605} ($V=16.3$\,mag, 
$\alpha_{2000}=05^{\mathrm h}16^{\mathrm m}28^{\mathrm s}$, $\delta_{2000}=+26^{\mathrm d}08^{\mathrm m}38^{\mathrm s}$)
was announced by \citet{1989AJ.....98.2221G}.
Later, several multi- and single-site observations followed the discovery run. Data were collected in 1992, 2000 
and 2001, respectively, including the Whole Earth Telescope (WET, \citealt{1990ApJ...361..309N}) run in 2000 November (XCov20), 
when \mbox{KUV 05134+2605} was one of the secondary targets. In this paper, we re-analyse these datasets. The log 
of observations and the reduction process applied in these cases are summarized in \citet{2003MNRAS.340.1031H}. 

To complement the earlier measurements, we collected data on \mbox{KUV 05134+2605} 
on 11 nights in the 2007/2008 observing season with the 1-m Ritchey-Chr\'etien-Coud\'e telescope at Piszk\'estet\H{o} 
mountain station of Konkoly Observatory. We continued the observations in 2011 for 3 nights. The measurements were made 
with a Princeton Instruments Vers\-Array:1300B back-illuminated CCD camera in white light and with 30\,s integration times. 
Table~\ref{tabl:log} shows the journal of the Konkoly observations. We obtained $\approx 48$ and $7$ hours of data in 2007/2008 
and 2011, respectively.

\begin{table}
\caption{Log of observations of \mbox{KUV 05134+2605}, performed at Piszk\'estet\H{o}, Konkoly Observatory.}
\label{tabl:log}
\centering
\begin{tabular}{p{3.5mm}ccrr}
\hline\hline
Run & UT date & Start time & Points & Length\\
No. & & (BJD-2\,450\,000) & & (h)\\
\hline
01 & 2007 Nov 29 & 4434.434 & 308 & 3.33\\
02 & 2007 Dec 01 & 4435.582 & 120 & 1.08\\
03 & 2007 Dec 04 & 4439.379 & 521 & 4.72\\
04 & 2007 Dec 05 & 4440.341 & 765 & 7.97\\
05 & 2008 Jan 23 & 4489.196 & 506 & 4.58\\
06 & 2008 Feb 09 & 4506.266 & 375 & 3.43\\
07 & 2008 Feb 11 & 4508.260 & 497 & 4.55\\
08 & 2008 Feb 12 & 4509.244 & 590 & 5.57\\
09 & 2008 Feb 18 & 4515.276 & 347 & 3.12\\
10 & 2008 Feb 19 & 4516.238 & 363 & 4.20\\
11 & 2008 Feb 20 & 4517.227 & 544 & 5.11\\
& & & &\\
12 & 2011 Feb 03 & 5596.235 & 233 & 2.12\\
13 & 2011 Feb 06 & 5599.298 & 179 & 1.71\\
14 & 2011 Feb 08 & 5601.229 & 354 & 3.19\\
\multicolumn{2}{l}{Total:} & \multicolumn{2}{r}{4936+766} & 47.66+7.02\\
\hline
\end{tabular}
\end{table}

\begin{figure}
\begin{center}
\includegraphics[width=6cm]{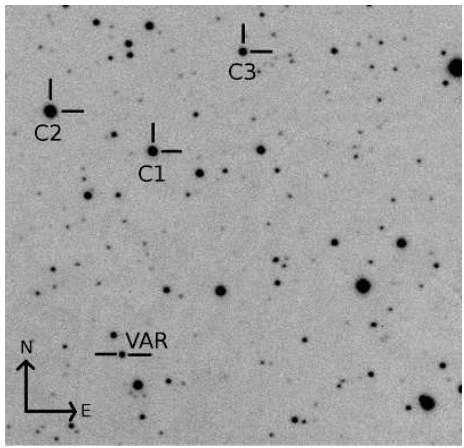}
\end{center}
\caption{One of the CCD frames obtained at Piszk\'estet\H{o}, Konkoly Observatory, with the variable and comparison 
stars marked. The field of view is $\approx7\arcmin\times7\arcmin$.}
\label{fig:comp}
\end{figure}

\begin{figure}
\begin{center}
\includegraphics[width=8.6cm]{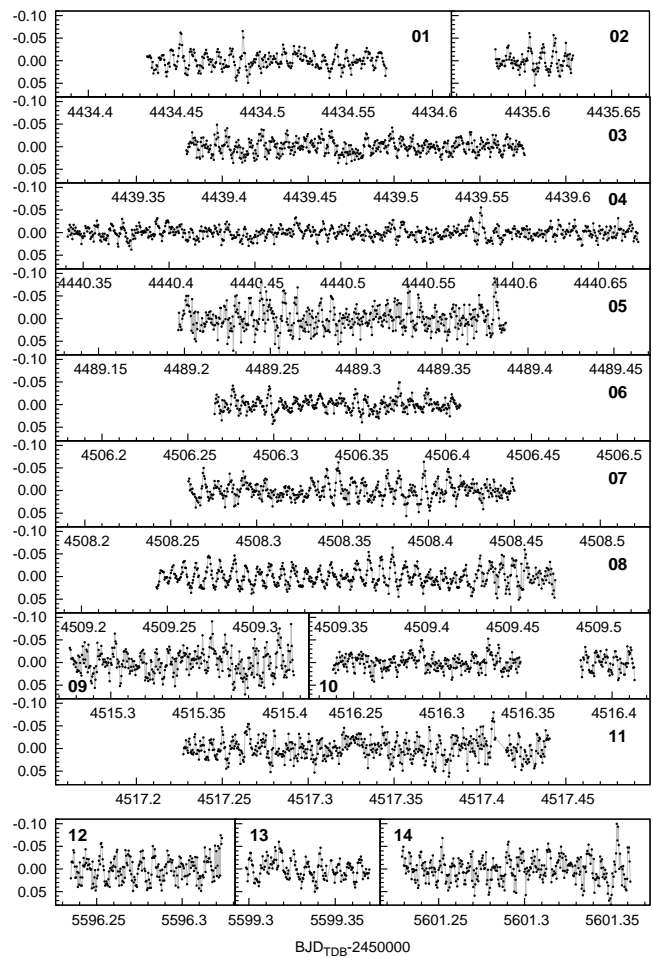}
\end{center}
\caption{Normalized differential light curves of the Konkoly runs. Data were collected in the 2007/2008 observing 
season (upper panels) and in 2011 (lower panels), respectively. For the better visibility of the pulsation cycles, 
we connected the points with grey lines. Boldface numbers in the corners correspond to the run numbers in 
Table~\ref{tabl:log}.}
\label{fig:lc}
\end{figure}

We followed the standard reduction procedure of the raw data frames: we applied bias, dark and flat corrections, and performed 
aperture photometry of the variable and comparison stars, using standard \textsc{iraf}\footnote{\textsc{iraf} is distributed 
by the National Optical Astronomy Observatories, which are operated by the Association of Universities for Research in Astronomy, 
Inc., under cooperative agreement with the National Science Foundation.} routines. We converted the JD times to Barycentric 
Julian Dates in Barycentric Dynamical Time ($\mathrm{BJD_{TDB}}$) using the applet of 
\citet{2010PASP..122..935E}\footnote{http://astroutils.astronomy.ohio-state.edu/time/utc2bjd.html}. After the aperture photometry, 
we checked whether the light curves of the possible comparison stars are free from any instrumental effects or variability. 
Finally, we selected three stars and used their average brightness as a comparison for the differential photometry. 
Fig.~\ref{fig:comp} shows the variable and the selected comparisons in the CCD field. We corrected for the atmospheric extinction 
and for the instrumental trends by applying low-order polynomial fits to the resulting light curves. This affected the 
low-frequency region in the Fourier Transforms (FTs) below $350\,\mu$Hz. Therefore we did not investigate this frequency domain.
Fig.~\ref{fig:lc} presents the light curves from the 2007/2008 and 2011 seasons, respectively. The Rayleigh frequency 
resolutions ($1/\Delta T$) of the whole light curves for these two epochs are $0.1$ and $2.3\,\mu$Hz. 

The most important parameters of the individual datasets used in this study can be found in Table~\ref{tabl:years}. 
This way, we can compare the different runs quantitatively, which can also be useful in the interpretation of their frequency 
content.

\section{Frequency content of the datasets}

We performed Fourier analysis of the datasets using the program \textsc{MuFrAn} (Multi-Frequency Analyser, 
\citealt{1990KOTN....1....1K, 2004ESASP.559..396C}). This tool can handle gapped and unequally spaced data, has Fast Fourier 
and Discrete Fourier Transform abilities, and is also capable of performing linear and non-linear fits of Fourier components. 
The results were also checked with \textsc{Period04} \citep{2005CoAst.146...53L} and the photometry modules of \textsc{FAMIAS} 
\citep{2008CoAst.155...17Z}. 
 
As Table~\ref{tabl:years} shows, the long time-base of the data obtained in the 2007/2008 season (hereafter `Konkoly2007 data') 
and the WET data from 2000 (hereafter `WET2000 data') provide the best Rayleigh frequency resolution and have the lowest noise 
levels. We review the results of the Fourier analyses of these two datasets first. 

\begin{table}
\caption{Comparison of the different observing runs on \mbox{KUV 05134+2605}. `Length' is the total time span of a run, 
$f_{\mathrm{R}}$ denotes the corresponding Rayleigh frequency resolution, `Cov.' refers to the temporal coverage parameter, 
`Noise' is the average amplitude value in the $4050-5050\,\mu$Hz frequency region, which is not affected by the star's 
pulsation, and `Points' is the number of data points.}
\label{tabl:years}
\centering
\begin{tabular}{lrcrcr}
\hline\hline
Run & Length & $f_{\mathrm{R}}$ & Cov. & Noise & Points\\
 & (d) & ($\mu$Hz) & (\%) & (mmag) & \\
\hline
1988 & 5.99 & 1.9 & 7.5 & 1.2 & 1257\\
1992 & 1.16 & 9.9 & 34.0 & 1.0 & 3415\\
2000 Oct & 5.11 & 2.3 & 8.2 & 0.8 & 3163\\
WET2000 & 13.07 & 0.9 & 11.1 & 0.4 & 9516\\
2001 & 2.09 & 5.5 & 14.8 & 1.4 & 1147\\
Konkoly2007 & 83.01 & 0.1 & 2.4 & 0.3 & 4936\\
Konkoly2011 & 5.13 & 2.3 & 5.7 & 1.1 & 766\\
\hline
\end{tabular}
\end{table}

\subsection{Analysis of the WET2000 and Konkoly2007 data}
\label{sect:longs}

\begin{figure*}
\begin{center}
\includegraphics[height=14.2cm]{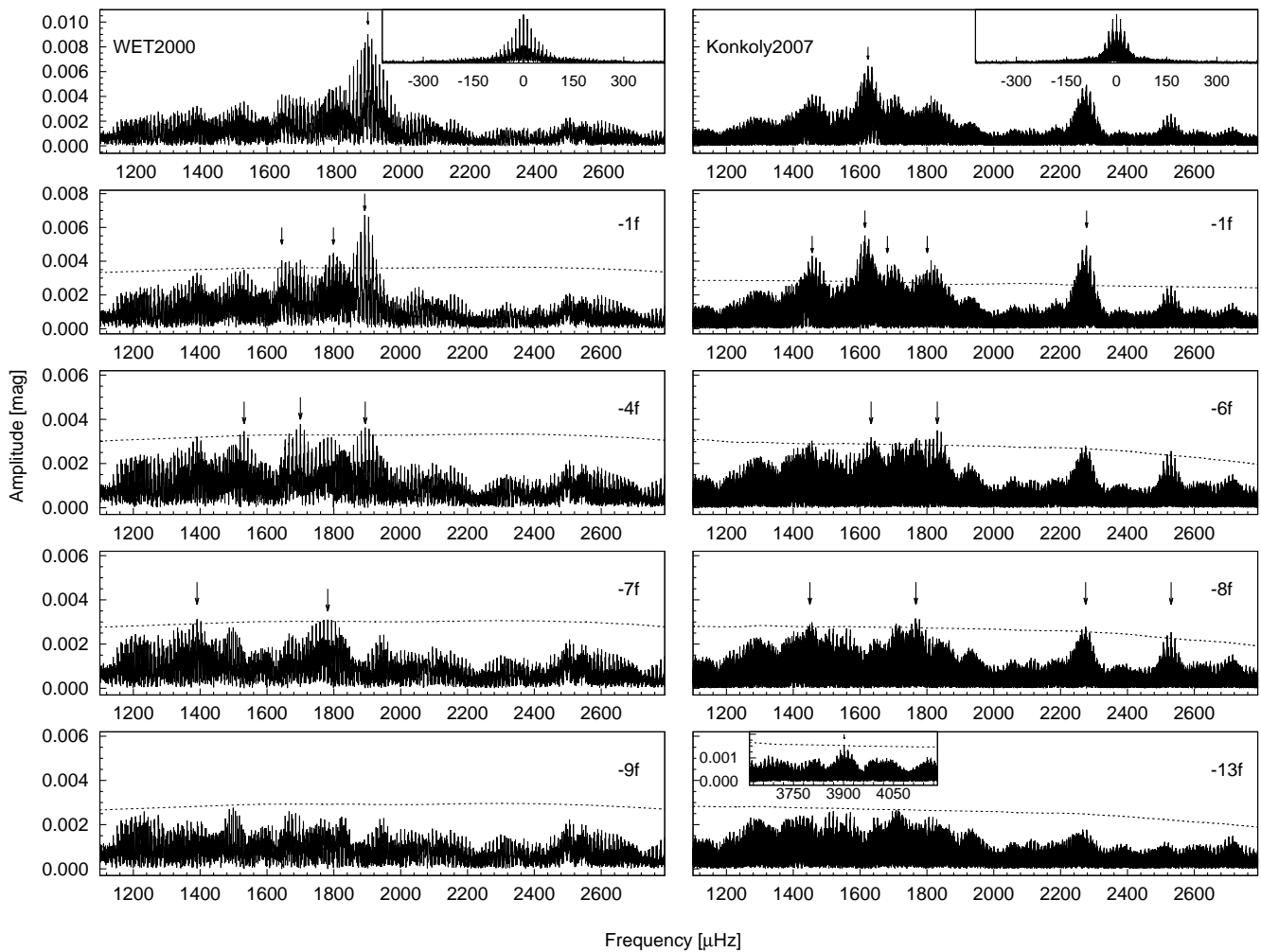}
\end{center}
\caption{Successive pre-whitening steps of the analyses of the WET2000 and Konkoly2007 datasets. The window functions are 
given in the inserts. Dashed lines denote the $4\langle A \rangle$ significance levels calculated by the moving average 
of the spectra.}
\label{fig:prewh}
\end{figure*}

Fig.~\ref{fig:prewh} presents the successive pre-whitening steps of the datasets. We checked not only the resulting 
amplitude, phase and frequency values, but the significance of the peaks in every fit. We took into account the peaks that 
reached the $4\langle A \rangle$ level (dashed lines), calculated by the moving average of radius $\sim1000\,\mu$Hz of 
the residual of the previous step's FT.

The choice of the frequencies showing the highest amplitudes was not automatic during the pre-whitening process, as the 
1\,d$^{-1}$ alias peaks have relatively large amplitudes (see the window functions inserted in Fig.~\ref{fig:prewh}). 
Thus, we fitted the data in different trials not only with the subsequent frequency having the highest amplitude, but also 
with its $\pm$1-2\,d$^{-1}$ 
aliases instead, and compared the resulting FTs and the associated light curve residual values (this represents the square root of the 
mean standard deviation between the fitted curve and the data). In the case of the Konkoly2007 dataset, the results of data
subsets' Fourier analyses also helped to decide which frequency we consider as real. For the frequency determination tests, see 
Sect.~\ref{sect:tests}.

The first columns of Table~\ref{table:prewh} summarize the pre-whitening results obtained on the WET2000 and 
Konkoly2007 data, showing the frequencies and amplitudes of the 9 and 13 peaks exceeding the $4\langle A \rangle$ 
significance level, respectively. 

\begin{table*}
\caption{Results of the frequency analyses of the datasets from different observing seasons. The largest amplitude 
frequency of a given dataset is marked with an asterisk. The abbreviation `K2011' refers to the new observations 
obtained at Konkoly Observatory in 2011. The numbers in parentheses are the uncertainties of the frequencies and amplitudes,
calculated by Monte Carlo simulations (see Sect.~\ref{sect:concl}).}
\label{table:prewh}
\centering
\begin{tabular}{rcrcrrr}
\hline\hline
\multicolumn{2}{c}{WET2000} & \multicolumn{2}{c}{Konkoly2007} & \multicolumn{3}{c}{Other measurements, $1988 - 2011$}\\
 \multicolumn{1}{c}{Frequency} & Amplitude & \multicolumn{1}{c}{Frequency} & Amplitude & \multicolumn{1}{c}{Frequency} & Amplitude & Dataset\\
 \multicolumn{1}{c}{($\mu$Hz)} & (mmag) & \multicolumn{1}{c}{($\mu$Hz)} & (mmag) & \multicolumn{1}{c}{($\mu$Hz)} & \multicolumn{1}{c}{(mmag)} & \\
 \hline
	  &	&	   & 	 & 1287.6(3.1) & 10.4(0.9) & 1988 \\
1390.6(2.8)  	& 3.2(0.3) & 	   & 	 & & & \\
	  &	&	   &	 & $^*$1414.7(0.1) & 26.6(1.0) & 1988 \\
	  &	& 	1449.7(3.3) & 3.3(0.3) & & & \\
	  &	& 	1456.6(4.1) & 4.4(0.4) &  1465.4(5.1) & 12.5(1.7) & 1992\\
	  &	& 	   & 	 & $^*$1472.5(1.2) & 17.7(1.1) & K2011 \\
	  &	&	   &	 & 1474.5(5.4) & 14.1(1.8) & 1992 \\
	  &	& 	   & 	 & 1504.5(0.1) & 17.3(1.1) & 1988 \\ 
1531.4(1.7)  	& 3.6(0.3)	&	   &	 & & & \\
	  &	& 	   & 	 & $^*$1549.7(4.9) & 12.5(0.8) & 1992 \\
	  &	& 	1614.5(1.6) & 6.2(0.4) & 1612.9(4.5) & 8.7(0.9) & 1992 \\
	  &	& 	$^*$1623.7(4.8) & 8.0(0.4) & & & \\
1644.2(4.5)  	& 4.4(0.3) & 1633.0(4.5) & 3.9(0.3) & & & \\
	  &	& 	1681.7(1.6) & 3.8(0.3) & & & \\
1699.9(2.3)  	& 4.2(0.3) & 	   &	 & & & \\
	  &	& 	1767.1(5.7) & 3.2(0.4) & $^*$1762.9(4.0) & 18.8(1.4) & 2001 \\
1781.7(2.3)  	& 3.5(0.3) & 	   &	 & & & \\
1798.6(2.3)  	& 5.3(0.4) &  1800.5(6.5) & 3.4(0.3) & & & \\ 
	  &	& 	1831.1(5.7) & 3.4(0.3) & & & \\
1892.3(0.1)  	& 5.7(0.3) & 	   &	 & 1892.2(5.3) & 13.7(0.7) & 2000 \\
1893.8(0.1) 	&  4.7(0.3) & &	 & & & \\
$^*$1901.9(0.1) & 8.4(0.5) & 	   &	 & $^*$1906.2(6.0) & 15.0(0.7) & 2000 \\
	  &	&	   &	 & 2023.7(7.7) & 8.4(1.1) & K2011 \\
	  &	& 2275.1(4.7) & 3.1(0.3)  & & & \\
	  &	& 2277.6(1.6) & 5.4(0.3) & & & \\
	  &	&	   &	 & 2504.7(6.6) & 5.1(0.8) & 1992 \\
	  &	& 2530.7(4.4) & 2.6(0.3) & & & \\
	  &	&	   &	 & 2806.6(6.3) & 10.5(0.9) & 1988 \\
	  &	&	   &	 & 2838.7(6.3) & 8.6(0.8)  & 1988 \\
	  &	& 3902.1(6.8) & 1.5(0.3) & & & \\
\hline
\end{tabular}
\end{table*}

In the WET2000 dataset, we detect two closely spaced frequencies as dominant ones (at $1901.9$ and $1892.3\,\mu$Hz) 
with a $9.6\,\mu$Hz frequency spacing (see the first two panels on the left of Fig.~\ref{fig:prewh}). The main pulsation 
frequency region is between $\approx 1200$ and $2000\,\mu$Hz.

The $1893.8\,\mu$Hz frequency is close to a higher-amplitude one ($\delta f = 1.5\,\mu$Hz, 
beat period: 7.5\,d). We considered and investigated in Sect.~\ref{sect:tests} the possibility that this is not a real 
pulsation frequency, but represents an example for an artificial peak which emerges as a result of temporal amplitude 
variations.

Similarly to the WET2000 data, two closely spaced frequencies dominate the Konkoly2007 dataset, but at $1623.7$ and 
$1614.5\,\mu$Hz. The dominant frequencies of the WET2000 data are below the detection limit. It is also 
important to note that an additional frequency can be detected at $1633.0\,\mu$Hz in the Konkoly2007 data, 
which completes the dominant doublet into an almost exactly equidistant triplet with spacings of $9.2$ and $9.3\,\mu$Hz 
(and corresponding beat periods of 1.3 and 1.2\,d), respectively. This is the only complete triplet structure detected in a 
single dataset.

In the Konkoly2007 data, besides the triplet, we find a doublet around $1450\,\mu$Hz, but with a different spacing: only 
$6.9\,\mu$Hz. In the $1700 - 1900\,\mu$Hz region, the number of peaks is rather high and the identification of the real 
pulsation peaks is difficult. We consider the three-frequency solution, in which one of the frequencies, at $1831.0\,\mu$Hz, 
can be detected also in a Konkoly2007 data subset (see Sect.~\ref{sect:tests}). The other two frequencies at $1767.1$ and 
$1800.5,\mu$Hz are the same within the uncertainties as the $1762.9$ and $1798.6\,\mu$Hz ones detected in the 2001 and WET2000 data,
respectively.

The $2275.1\,\mu$Hz frequency is also considerably close to a larger amplitude one at $2277.6\,\mu$Hz, similarly to
the case of the $1893.8\,\mu$Hz frequency of the WET2000 data. 

The $3902.1\,\mu$Hz frequency in the Konkoly2007 dataset is a combination ($f_{\mathrm 1} + f_{\mathrm 2}$) of the two 
large-amplitude ones at $1623.7$ and $2277.6\,\mu$Hz. We did not detect other combination frequencies neither in the 
WET2000 nor in the Konkoly2007 dataset.

In conclusion, we find a common frequency of the WET2000 and Konkoly2007 datasets at $1800\,\mu$Hz. The frequencies 
at $1633.0\,\mu$Hz (2007) and $1644.2$ (2000) are also common, if we consider that these are most likely 1\,d$^{-1}$ 
($11.574\,\mu$Hz) aliases. 

The peaks found at $1681.7$ and $1767.1\,\mu$Hz (2007) are also slightly different from the ones detected at 
$1699.9$ and $1781.7\,\mu$Hz (2000). Different scenarios can explain the emergence of these frequency pairs (alias ambiguities, 
rotational splitting, the mixing of dipole and quadrupole modes). We discuss them in Sect.~\ref{sect:26list}.

\subsection{Analysis of the short datasets obtained between 1988 and 2011}
\label{sect:analysis}

Table~\ref{table:prewh} summarizes the results of the Fourier analysis of the short light curves, too. The largest number of 
frequencies are detected in the 1988 and 1992 data (5-5 frequencies). In 2000, 2001, and 2011, two, one, and two frequencies
exceeded the $4\langle A \rangle$ significance level, respectively. The dominant frequency, around $1901\,\mu$Hz,
did not change from 2000 October to 2000 November-December (WET2000), while it was different in all the other observing seasons. 

\textit{1988}: The $1287.6$ and the $1504.5\,\mu$Hz frequencies are candidates for new pulsation modes. The $1414.7\,\mu$Hz 
frequency is close to the 2\,d$^{-1}$ alias of the $1390.6\,\mu$Hz peak of the WET2000 dataset, but it could still be an
independent frequency. The two frequencies above $2800\,\mu$Hz can be interpreted as combination terms, considering the alias 
ambiguities. 

\textit{1992}: In the 1992 dataset, we detect a doublet at $1465.4$ and $1474.5\,\mu$Hz with $9.1\,\mu$Hz separation.
We note that this separation is slightly lower than the Rayleigh frequency resolution of the dataset ($9.9\,\mu$Hz). We checked
the reliability of this doublet in the course of the frequency determination tests, see Sects.~\ref{sect:tests} 
and \ref{sect:26list}.
The $1612.9\,\mu$Hz frequency corresponds to the $1614.5\,\mu$Hz one from 2007. The $2504.7\,\mu$Hz 
frequency might be a -2\,d$^{-1}$ alias of the $2530.7\,\mu$Hz one detected also in 2007, or might represent a new pulsation 
frequency.

\textit{2000}: The two significant frequencies obtained in 2000 October correspond to the two largest amplitude ones detected 
in the WET2000 dataset. 

\textit{2001}: The only significant frequency of the 2001 data at $1762.9\,\mu$Hz is the same within the uncertainties as the 
$1767.1\,\mu$Hz one observed in 2007, so this confirms the choice of this latter during the pre-whitening process. 

\textit{2011}: The $2023.7\,\mu$Hz peak is a promising candidate for a new pulsation mode. The frequency at $1472\,\mu$Hz 
was found in the 1992 data.

\subsection{Frequency determination tests}
\label{sect:tests}

We performed several different tests based both on observational and synthetic data to investigate how reliable are the 
frequency solutions presented in Table~\ref{table:prewh}. These tests also help to calculate more realistic uncertainties for the 
frequency values than the simple formal uncertainties derived from the light curve fits.

\textit{Analysis of data subsets}: The Konkoly2007 dataset provides the longest time base, but it has also the 
worst temporal coverage. We selected four data subsets for independent analysis: subset 1 (runs No.\,3 and 4 in 
Table~\ref{tabl:log}), subset 2 (runs No.\,$1-4$), subset 3 (runs No.\,$6-8$) and subset 4 (runs No.\,$9-11$).
Comparing to the result of the Fourier analysis of the whole dataset (Table~\ref{table:prewh}), we found five frequencies
being significant also in the subset data. Four of them (at $1456$, $1624$, $1633$ and $1831\,\mu$Hz) could be detected only once, 
but we found the $2277\,\mu$Hz frequency in every subset.

\textit{Analysis of synthetic datasets}: We performed this test separately for every dataset: the 1988, 1992, 
2000 Oct, WET2000, 2001, Konkoly2007, Konkoly2011, and also the subset $1-4$ data. In this step, we generated 
at least 50 synthetic light curves with the same timings as the original data and with Gaussian 
random noise added, using the frequencies, amplitudes and phases of the corresponding light curve solution presented in Table~\ref{table:prewh}. 
The applied noise level was calculated according to the residual scatter of the given light curve solution. 
We carried out independent and automated frequency analysis of each synthetic dataset with the program \textsc{SigSpec} 
\citep{2007A&A...467.1353R}. This way we could eliminate the subjective personal factor e.g. in the selection of the 
starting frequency. We checked finally in what percentage the program found the input frequencies instead of their aliases. 
The success rates we obtained for the different frequencies are between 34 and 100 perc cent, but above 70 per cent in most cases.

\textit{Analysis of synthetic datasets with amplitude variations}: The presence of temporal variations in pulsating 
white dwarfs are well-known, and comparing the FTs of different epochs' datasets, amplitude variations 
of \mbox{KUV 05134+2605} are obvious. However, the Fourier analysis method we apply for the pre-whitening assumes constant 
amplitudes, frequencies and phases for the pulsation. Any deviations from this assumption may result in the emergence of 
additional peaks in the Fourier spectrum.

We divided the Konkoly2007 dataset into five subsets: 
subset \textsc{a}, \textsc{b}, \textsc{c}, \textsc{d} and \textsc{e} (runs No.\,1--2, 3--4, 5, 6--8 and 9-11), and generated 
synthetic light curves for the times of these subsets. We fixed the Konkoly2007 frequencies and phases according to the ones 
determined by the whole dataset, but let the amplitudes vary from subset to subset for every frequency. 
We omitted only the combination term at $3902.1\,\mu$Hz and the $2275.1\,\mu$Hz frequency from the test. This
latter may be an artefact resulting from temporal variations in the pulsation, and this way we checked, whether amplitude 
changes can really cause such closely spaced peaks. We chose the amplitudes applied to represent both slight and major
variations from subset to subset, based on the observed values.
Constructing this way a new, synthetic Konkoly2007 dataset with variable amplitudes, we then created 50 of them adding Gaussian 
random noise as in the case of the fixed-amplitude tests, and analysed them with \textsc{SigSpec}. The program found the original 
frequencies with 22--98 per cent success rates, and 6 out of the 11 frequencies reached or exceeded the 70 per cent rate.

The other finding is that amplitude variations can cause indeed the emergence of closely spaced, and low amplitude peaks in 
the Fourier transforms. For example, the doublet at $2277.6$ and $2278.5\,\mu$Hz showing $3.9$ and $2.8$\,mmag 
amplitudes and $0.9\,\mu$Hz frequency separation in one of the test datasets. We note that similarly closely spaced peaks 
were detected, e.g., in the case of the ZZ Ceti stars \mbox{\object{KUV 02464+3239}} \citep{2009MNRAS.399.1954B} and 
\object{EC14012-1446} \citep{2012ApJ...751...91P}. Both stars show variable amplitudes on a time-scale of weeks.

\subsubsection{Conclusions}
\label{sect:concl}

The main goal of these tests was to investigate the robustness of the light curve solutions presented in 
Table~\ref{table:prewh}. Based on the results, we constructed the list of frequencies considered real (see 
Table~\ref{table:finalfrek}) applying the following rules: (1) in the case of the Konkoly2007 frequencies, we considered real 
a frequency if it was detected at least 75 per cent success rate at least in 2 test cases (considering the analyses of data 
subsets and the whole dataset with fixed or variable amplitudes). (2) In the case of the other datasets, we considered real 
a frequency if was found at least 75 per cent success rate in the corresponding test data. (3) If two frequencies found to
be the same within the uncertainties, we chose the one having the lower uncertainty. (4) Finally, we do not consider real the $1893.8$ 
and $2275.1\,\mu$Hz frequencies, as based on the test with variable amplitude frequencies, we cannot exclude the possibility 
that these closely spaced peaks are artefacts from such variations.

However, these simple rules does not help to choose between the $1644.2$ and $1633.0\,\mu$Hz 1\,d$^{-1}$ alias frequencies. 
We included the $1633.0\,\mu$Hz one in the list of Table~\ref{table:finalfrek}, but with the additional note that its
$+1$\,d$^{-1}$ alias solution is also acceptable.

The uncertainties presented both in Tables~\ref{table:prewh} and \ref{table:finalfrek} are calculated by the analyses of the 
test light curves. In calculations with Monte Carlo simulations, we determine the standard deviations of the 
input frequencies derived from a large number of test datasets. In this case, we calculated the uncertainties 
including the cases when the program found not the original frequency, but one of its alias peaks. This way we took into account 
the effect of alias ambiguities in the uncertainties.

\subsection{Investigation of the set of frequencies}
\label{sect:26list}

In total, 22 frequencies were considered as the most probable set of frequencies after the tests described in Sect.~\ref{sect:tests}. 
Table~\ref{table:finalfrek} lists these and also summarizes the frequency spacings. Considering these spacing values, we can 
form groups of frequencies, in which the separation of the consecutive members are less than $26\,\mu$Hz 
($\sim2\times 1$\,d$^{-1}$, taking the uncertainties  of frequencies into account). These groups are also indicated in 
Table~\ref{table:finalfrek}. The frequency separation was chosen to be $\sim2\times 1$\,d$^{-1}$ in order to ensure
that even if one or both frequencies suffer from $1$\,d$^{-1}$ alias ambiguities, they remain in the same group. 

\begin{table}
\caption{Frequencies and periods of the accepted pulsation frequencies of \mbox{KUV 05134+2605}. The third column shows the 
frequency differences of closely spaced peaks. The last two columns summarize the results of the preliminary mode 
identification based on rotational multiplets and period spacing investigations (see Sect.~\ref{sect:pp}). The frequency uncertainties 
are the same as in Table~\ref{table:prewh}.}
\label{table:finalfrek}
\centering
\begin{tabular}{lrrrr}
\hline\hline
\multicolumn{1}{c}{Frequency} & \multicolumn{1}{c}{Period} & \multicolumn{1}{c}{$\delta f$} & \multicolumn{1}{c}{$\ell$} & \multicolumn{1}{c}{$m$}\\
\multicolumn{1}{c}{($\mu$Hz)} & \multicolumn{1}{c}{(s)} & ($\mu$Hz) & & \\
\hline
1287.6\,$\pm$3.1 & $^+$776.6\,$\pm$1.9 & & 1 & \\
\\
1390.6\,$\pm$2.8 & $^+$719.1\,$\pm$1.4 & 24.1 & 1 & \\
1414.7\,$\pm$0.1 & $^+$706.8\,$\pm$0.1 & & 2,1 & \\
\\
1456.6\,$\pm$4.1 & 686.6\,$\pm$1.9 & 15.9 & 1,2 & \\
1472.5\,$\pm$1.2 & $^+$679.1\,$\pm$0.6 & & 1,2 & \\
\\ 
1504.5\,$\pm$0.1 & $^+$664.7\,$\pm$0.1 & & 2 & \\
\\
1531.4\,$\pm$1.7 & 653.0\,$\pm$0.7 & 18.3 & 1 & -1\\
1549.7\,$\pm$4.9 & $^+$645.3\,$\pm$2.0 & & 1 & +1\\
\\
1614.5\,$\pm$1.6 & 619.4\,$\pm$0.6 & 9.2 & 1 & -1\\
1623.7\,$\pm$4.8 & 615.9\,$\pm$1.8 & 9.3 & 1 & 0\\
1633.0\,$\pm$4.5\tablefootmark{a} & 612.4\,$\pm$1.7 & & 1 & +1\\
\\
1681.7\,$\pm$1.6 & 594.6\,$\pm$0.6 & 18.2 & 1 & -1\\
1699.9\,$\pm$2.3 & 588.3\,$\pm$0.8 & & 1 & +1\\
\\
1762.9\,$\pm$4.0 & $^+$567.2\,$\pm$1.3 & 18.8 & 2 & \\
1781.7\,$\pm$2.3 & 561.2\,$\pm$0.7 & 16.9 & 1 & -1\\
1798.6\,$\pm$2.3 & $^+$556.0\,$\pm$0.7 & & 1 & +1\\
\\
1831.1\,$\pm$5.7 & 546.1\,$\pm$1.7 & & 2 & \\
\\
1892.3\,$\pm$0.1 & 528.5\,$\pm$0.1 & 9.6 & 1 & -1,0\\
1901.9\,$\pm$0.1 & $^+$525.8\,$\pm$0.1 & & 1 & 0,+1\\
\\
2277.6\,$\pm$1.6 & 439.1\,$\pm$0.3 & & 2,1 & \\
\\
2504.7\,$\pm$6.6 & 399.2\,$\pm$1.1 & 26.0 & 1 & \\
2530.7\,$\pm$4.4 & 395.1\,$\pm$0.7 & & 2,1 & \\
\hline
\end{tabular}
\tablefoot{
Periods marked with plus symbol ($\,^+\,$) are reported previously by 
\citet{2003MNRAS.340.1031H} as dominant signals at different epochs.\\
\tablefoottext{a}{+1\,d$^{-1}$ solution ($1644.2\,\mu$Hz) is also possible by the WET2000 dataset.}
}
\end{table}

The first conspicuous thing is the large number of excited modes. The grouping of frequencies suggests that at least 
12 normal modes can be detected by the ensemble treatment of the different datasets. This means that \mbox{KUV 05134+2605} 
has become one of the `rich' white dwarf pulsators known. We know only a few of them \citep{2009AIPC.1170..621B, 2012ASPC..462..164B}, 
and more modes give stronger constraints for the asteroseismic investigations. Note that when the modes are asymptotic, adding 
one or two new frequencies to the list of equidistant modes brings limited new information. This is not the case of 
\mbox{KUV 05134+2605} (see Table~\ref{table:finalfrek}). Consequently, \mbox{KUV 05134+2605} has a particular importance among 
the white dwarf variables, especially the V777 Her-type ones. 

We have already mentioned in Sects.~\ref{sect:longs} and \ref{sect:analysis} the doublets and the triplet showing 
$\approx 9\,\mu$Hz frequency spacings. The 1992 dataset's doublet failed in the frequency determination test, 
as we could get back both peaks \textit{in the same dataset} only in 26 per cent of the test cases, i.e., with a low success rate. 
However, this is not surprising because their separation is close to the dataset's Rayleigh frequency resolution. 
The $1892 - 1901\,\mu$Hz doublet found by the WET2000 data seems to be a robust finding, as the test resulted 100 per cent success 
rate for the determination of both components. 
Considering the triplet at $1623\,\mu$Hz in the Konkoly2007 dataset, we could detect all three components in 74 per cent of the fixed-amplitude 
cases. We also find two of the components in the subset 3 data, with 82 per cent doublet-finding rate. The recurring emergence 
of the $\approx 9\,\mu$Hz spacing in different datasets (epochs) with different window functions strongly suggests that this is 
not the result of the 1\,d$^{-1}$ alias problem, but these frequency components have pulsational origin.

It is also conspicuous that frequencies were found not only with $\approx 9\,\mu$Hz spacings, but also with its double. 
However, we must be cautious with the interpretation of these spacings; as the doublet components originate from different 
datasets, there is a risk that we see outcome of the frequency determination uncertainties and the 1\,d$^{-1}$ alias problem.
If these frequencies are real, a possible explanations is that the frequencies showing $\approx 18\,\mu$Hz spacings 
are $m\pm 1$ peaks of the same triplet.

Table~\ref{table:finalfrek} also shows doublets with $\approx 16$ and $26\,\mu$Hz spacings. We discuss them in 
Sects.~\ref{sect:rot} and \ref{sect:pp}.

\subsubsection{Stellar rotation}
\label{sect:rot}

Because of the better visibility of dipole ($\ell=1$) modes over quadrupole ($\ell=2$) ones (see e.g. \citealt{2008MNRAS.385..430C} 
and references therein), we expect that most of the detected frequencies are $\ell=1$. We see that there is a recurring
$\approx 9\,\mu$Hz spacing in our datasets, and the most robust finding is the $\delta f = 9.6\,\mu$Hz doublet in the WET2000
data. Supposing that the frequency components are high-overtone ($k\gg1$) $\ell=1$, 
$m=-1,0$ or $m=0,1$ modes, the rotation period of the star can be calculated by the asymptotic relation: 
$\delta f_{k,\ell,m} = \delta m (1-C_{k,\ell}) \Omega$, where the coefficient \mbox{$C_{k,\ell} \simeq 1/\ell(\ell+1)$} and 
$\Omega$ is the rotation frequency. Thus, the $9.6\,\mu$Hz splitting value implies a 0.6\,d rotation period for 
\mbox{KUV 05134+2605}.

According to theory \citep{1991ApJ...378..326W}, the ratio of the rotational splitting for $\ell=1$ and $\ell=2$ $g$-modes is 
$\delta f_{\ell=1} / \delta f_{\ell=2} = 0.60$ in the asymptotic limit, as $\delta f \sim [1-1/\ell(\ell+1)]$. We find 
two closely spaced frequencies having $\delta f = 16.9\,\mu$Hz in the WET2000 dataset. Assuming that these are components 
of rotationally split $\ell=2$ modes, the corresponding splitting ratio is ($9.6\pm0.1)/(16.9\pm3.3)=0.57\pm0.11$. This is the same
within the errors as the theoretical 0.60 value. However, these two frequencies still can be $m=-1,+1$, $\ell=1$ 
components, too. Note that the $15.9\,\mu$Hz separation of the $1456$ and $1472\,\mu$Hz 
frequencies also could be interpreted as rotational splitting of $\ell=2$ modes ($9.6/15.9=0.60$).

\subsubsection{Period spacings}
\label{sect:pp}

\begin{figure*}
\begin{center}
\includegraphics[width=17cm]{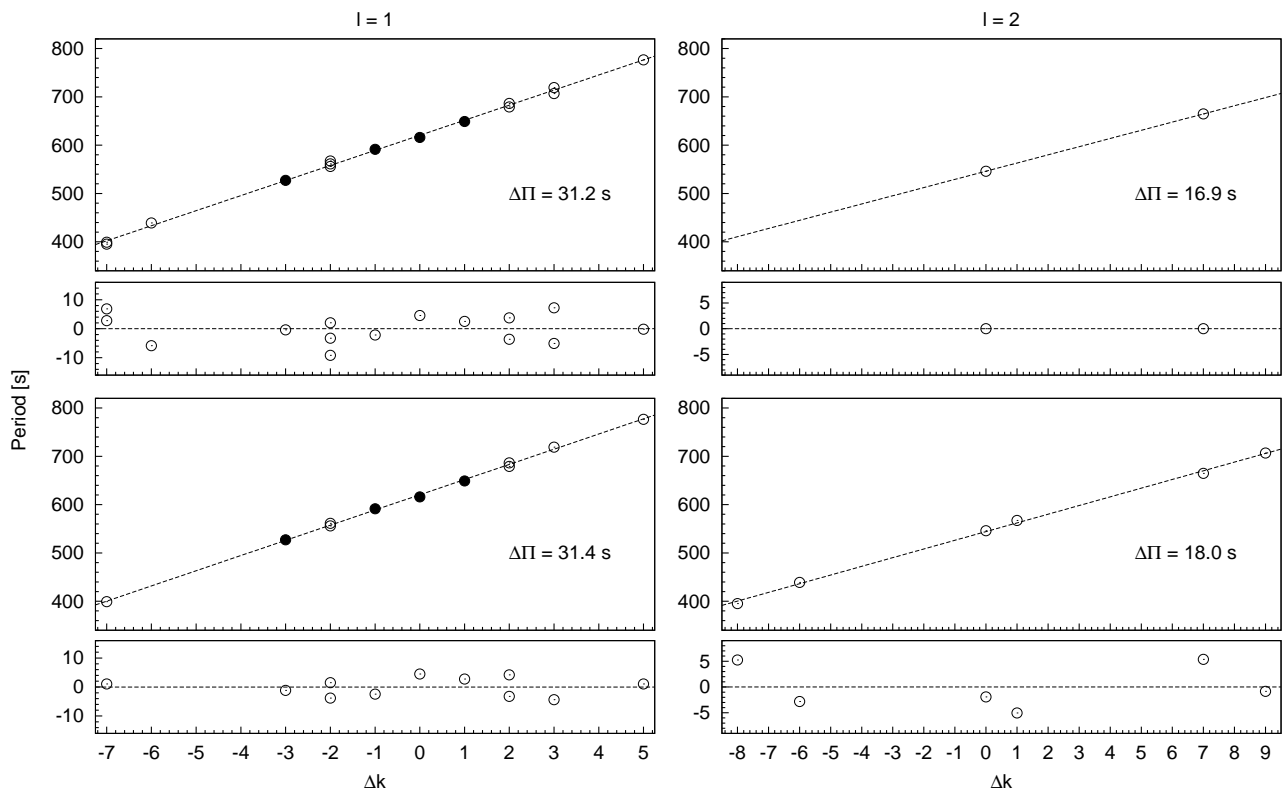}
\end{center}
\caption{Periods of Table~\ref{table:finalfrek} plotted in the $\Delta k$ -- period planes. Filled circles denote average 
periods by the $9\,\mu$Hz spacings. The panels on 
the left and right show the $\ell=1$ and 2 solutions and their linear least-squares fits, respectively. The residuals of 
the fits are also presented. Top and bottom rows display different solutions by changing some periods' presumed $\ell$ values. 
The resulting period spacings ($\Delta \Pi_{\ell=1,2}$) are denoted.}
\label{fig:pfits}
\end{figure*}

For $g$-modes with high radial order $k$ (long periods), the separation of consecutive periods ($\Delta k= 1$) becomes nearly
constant at a value given by the asymptotic theory of non-radial stellar pulsations. Specifically, the asymptotic period spacing 
\citep{1990ApJS...72..335T} is given by:

\begin{equation} 
\Delta \Pi_{\ell}^{\rm a}= \Pi_0 / \sqrt{\ell(\ell+1)}, 
\label{aps}
\end{equation}
\noindent where
\begin{equation}
\label{asympeq}
\Pi_0= 2 \pi^2 \left[ \int_{r_1}^{r_2} \frac{N}{r} dr\right]^{-1},
\end{equation}

\noindent $N$ being the Brunt-V\"ais\"al\"a frequency. 
In principle, one can compare the asymptotic period spacing
computed from a grid of models with different masses and effective
temperatures with the mean period spacing exhibited by the star, and
then infer the value of the stellar mass. This method has been applied
in numerous studies of pulsating PG 1159 stars (see, for instance, 
\citealt{2007A&A...461.1095C, 2007A&A...475..619C, 2008A&A...478..869C, 
2009A&A...499..257C}, and references therein). For the
method to be valid, the periods exhibited by the pulsating star must
be associated with high order $g$-modes, that is, the star must be
within the asymptotic regime of pulsations.
In addition, the star should be chemically homogeneous 
for Eq.~(\ref{aps}) to be strictly valid. 
However, the interior of DB white dwarf stars are
supposed to be chemically stratified and characterized by strong
chemical gradients built up during the progenitor star life. So, the
direct application of the asymptotic period spacing to infer the
effective temperature and stellar mass of \mbox{KUV 05134+2605} is 
questionable. A better way to compare our models to the
period spacing of the observed pulsation spectrum is to calculate the
average of the computed period spacings, using:

\begin{equation}
\label{avgdp}
\overline{\Delta \Pi}(M_*, T_{\rm eff})= \frac{1}{(n-1)} 
\sum_{k}^{n-1} \Delta \Pi_k, 
\end{equation}

\noindent where $\Delta \Pi_k$ is the `forward' period spacing
defined as $\Delta \Pi_k= \Pi_{k+1}-\Pi_k$, and $n$ is the number of
computed periods laying in the range of the observed periods. The
theoretical period spacing of the models as computed 
through Eqs.~(\ref{aps}) and (\ref{avgdp})
share the same general trends (that is, the same dependence on $M_*$
and $T_{\rm eff}$), although $\Delta \Pi_{\ell}^{\rm a}$ is usually
somewhat higher than $\overline{\Delta \Pi}$. 

As a preliminary mode identification, we attempted to determine the $\ell$ values of the observed 
frequencies using not only the effect of rotational splitting, but the period spacing values derived. 
The period values of the frequencies are given in Table~\ref{table:finalfrek}.

Assuming that the frequencies showing $\delta f \approx 9\,\mu$Hz spacings or its double are $\ell=1$, we
created an initial set of the (supposed) $\ell=1$ modes: \textit{527.2, 591.4, 615.9}, and \textit{649.1}\,s. These are the 
averages of the triplet and doublet peaks observed. This way we can avoid any assumption on the actual $m$ values.
The period differences in this case are 64.3 ($2\times32.2$), 24.5 and 33.2\,s, respectively.
We did not use the average of the peaks at 560\,s, but investigated the three frequencies separately.

The effective temperature and surface gravity parameters of \mbox{KUV 05134+2605} are $T_{\mathrm{eff}}=24\,700\pm1300$\,K 
and $\mathrm{log\,} g=8.21\pm0.06$ ($M_*=0.73\pm0.04\,M_\odot$), from model atmosphere fits to a high signal-to-noise 
optical spectrum of the star \citep{2011ApJ...737...28B}. \citet{2011ApJ...737...28B} also found that \mbox{KUV 05134+2605} 
belongs to the DBA spectral class, which means that it shows not only helium, but hydrogen lines.

Theoretical calculations show that period spacings of $\ell=1$ modes of 0.7 and 0.8\,$M_*$ model stars can vary between 
$\approx 25-40$\,s, depending on mode trapping \citep{1993ApJ...406..661B}. It is also known that the mean period spacing 
decreases with increasing stellar mass. Considering Fig. 8 of \citet{1993ApJ...406..661B}, the minima in the period spacing 
diagrams of 0.7 and 0.8\,$M_*$ model stars are $\approx 30$ and 25\,s for the $\ell=1$ modes, respectively. Trapped modes 
have lower kinetic energies and minima in period spacing diagrams correspond to minima in kinetic energies 
\citep{1991ApJS...75..463B}. Assuming that there is a missing overtone in our period list, the period spacing values are 
$\approx 30$\,s, except one case with the lower value of 25\,s. These are acceptable values of period spacings for this 
relatively high-mass variable.

\citet{2013MNRAS.429.1585F} demonstrated in the case of the ZZ Ceti star \object{HS 0507+0434B}, that the mean 
period spacing value can also be determined by the linear least-squares fit of the $m=0$ modes plotted in the 
$\Delta k$ -- period plane. Assuming that most of the observed frequencies correspond to $\ell=1$ modes, we fitted almost 
all of the periods on the top left panel of Fig.~\ref{fig:pfits}, excluding only the periods at 664 and 546\,s; 
because of the high frequency density in these regions, it seems improbable that they are all $\ell=1$. This solution 
reflects the expectation that the period differences should be higher and lower than 25\,s for the $\ell=1$ and 2 modes, 
respectively. The formula used for the fits is $P_k = P_\mathrm{0} + \Delta P \times \Delta k$ \citep{2013MNRAS.429.1585F}. 
In the case of the supposed $\ell=1$ modes, we selected the 615.9\,s mode as $k_0$. 

The top panels of Fig.~\ref{fig:pfits} show the corresponding $\ell=1$ and $2$ fits and their residuals. Filled circles 
denote the four average periods of the presumed $\ell=1$ modes by the $9\,\mu$Hz spacings. The average $\ell=1$ period 
spacing determined this way is $\Delta \Pi_{l=1} = 31.2\pm0.4$\,s. We present a possible fit to the two $\ell=2$ periods 
($\Delta \Pi_{l=2} = 16.9$\,s) for comparison with the next fit's result. 

As we see some outliers in the $\ell=1$ fit, we assumed that their previously attributed $\ell$ values are not correct, 
and in the next step, we moved them to the $\ell=2$ group in an attempt to optimize our fit. The bottom panels of 
Fig.~\ref{fig:pfits} show the results of this solution. The corresponding $\Delta \Pi$ values for the $\ell=1$ and 2 modes 
are $31.4\pm0.3$ and $18.0\pm0.3$\,s, respectively. The results show that with the change of the $\ell$ values of 
some periods, the average period spacings change little. Note that this method alone cannot be used for mode identification, 
but strongly suggests that even if there are frequencies in our list which are not correctly determined because of the 
alias ambiguities, the frequencies are not randomly distributed, and especially in the case of the presumed $\ell=1$ ones, 
points to a definite mean spacing value.

According to theoretical calculations (e.g. \citealt{2011MNRAS.414.2885R}) the asymptotic ratio of the period spacings 
of $\ell=1$ and 2 modes is $\Delta \Pi_{l=1} / \Delta \Pi_{l=2} = \sqrt{3} = 1.73$. For the second fits
presented above, this ratio is $(31.4\pm0.3)/(18.0\pm0.3) = 1.74\pm0.03$. That is, the result agrees 
with the theoretical prediction within the uncertainty.

In the case of the frequency pairs, we can conclude that (1) the 719 and 706\,s periods, if both are real pulsation frequencies 
and not aliases, cannot have the same $\ell$, and the 706\,s one may be the $\ell=2$. (2) The $686 - 679$\,s peaks either 
can be rotationally split or 1\,d$^{-1}$ alias $\ell=1$ or $2$ frequencies.
(3) According to the period fits, the 561 and 556\,s periods may be rotationally split $\ell=1$ frequencies ($m=-1,+1$) and 
not $\ell=2$ ones.
In this case, the 567\,s period is an $\ell=2$, and its $18.8\,\mu$Hz frequency separation with the 561\,s peak is a coincidence, 
and have nothing to do with the supposed $9\,\mu$Hz rotational splitting. (4) The 399\,s member of the $395-399$\,s pair fits 
better into the series of $\ell=1$ modes. The last two columns of Table~\ref{table:finalfrek} show the presumed 
$\ell$ and $m$ values determined by the supposed rotational multiplets and the period spacing investigations.

\subsubsection{Tests on period spacings}
\label{sect:testspp}

All of our efforts aim to know more about the stellar structure, specially the mass of the observed object. As we 
mentioned above, the mean period spacing is an indicator of the stellar mass, and applying linear fits to the observed 
periods, we determined $31.4$\,s value for the presumed $\ell = 1$ modes. Considering the 1\,d$^{-1}$ alias ambiguities, 
the reliability of this value must be investigated. We performed several tests to check our solution.

\begin{figure}
\centering
\includegraphics[width=9.0cm]{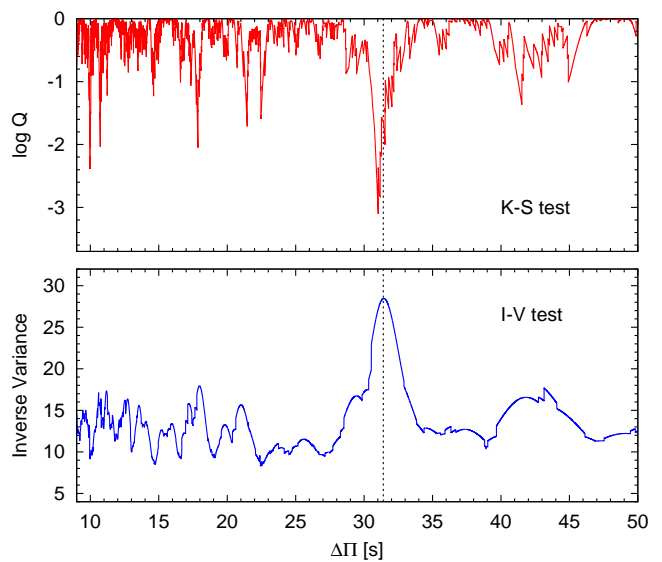}
\caption{Characteristic spacings of the periods listed in Table~\ref{table:finalfrek}, applying the Kolmogorov-Smirnov 
(K-S) and the Inverse Variance (I-V) significance tests. The vertical dotted line denotes the 31.4\,s 
spacing value obtained in Sect.~\ref{sect:pp}.} 
\label{fig:tests1}
\end{figure}

We searched for a characteristic period spacing between the periods listed in Table~\ref{table:finalfrek} by using the 
Kolmogorov-Smirnov (K-S) \citep{1988IAUS..123..329K} and the Inverse Variance (I-V) \citep{1994MNRAS.270..222O} significance 
tests. In the K-S test, the quantity $Q$ is defined as the probability that the observed periods are randomly distributed. 
Thus, any characteristic period spacings present in the period spectrum should appear as a minimum in $Q$. In the I-V test, 
a maximum of the inverse variance will indicate the presence of a constant period spacing. Taking all the periods of 
Table~\ref{table:finalfrek}, thus without any assumptions based on frequency or period spacings, both methods 
strongly point to the existence of 
a characteristic period spacing of about 31\,s (see Fig.~\ref{fig:tests1}). The minimum in the K-S test is at $31.0$\,s, 
while the maximum in the I-V plot is at $31.4$\,s, respectively. These values are in very good agreement with the mean period 
spacings found by linear fits for $\ell=1$ modes.

Another well-known method to find a characteristic spacing value is the Fourier analysis of the periods themselves 
(see e.g. \citealt{1997MNRAS.286..303H}). At first, we performed the Fourier analysis of the periods listed in 
Table~\ref{table:finalfrek} and found a clear maximum at $31.4$\,s. This also supports the presence of the spacing 
derived by the linear fits, and the K-S and I-V tests. In the next step, we modified the period list used: we chose 
randomly 8 frequencies and (also randomly) added or subtracted $11.574\,\mu$Hz (1\,d$^{-1}$) to these values. That 
is, we changed almost 40 per cent of the periods. Afterwards, 
we generated a new period list with these modified frequencies, simulating the effect of the 1\,d$^{-1}$ aliasing, 
present in all datasets. We generated 20 such aliased period lists and performed their Fourier analyses. 
Applying this Monte Carlo test, we still find a common maximum around 31\,s in each simulation. The average of the 
maxima for the 20 different cases is $31.4\pm0.2$\,s. 

\begin{figure}
\centering
\includegraphics[width=7.35cm]{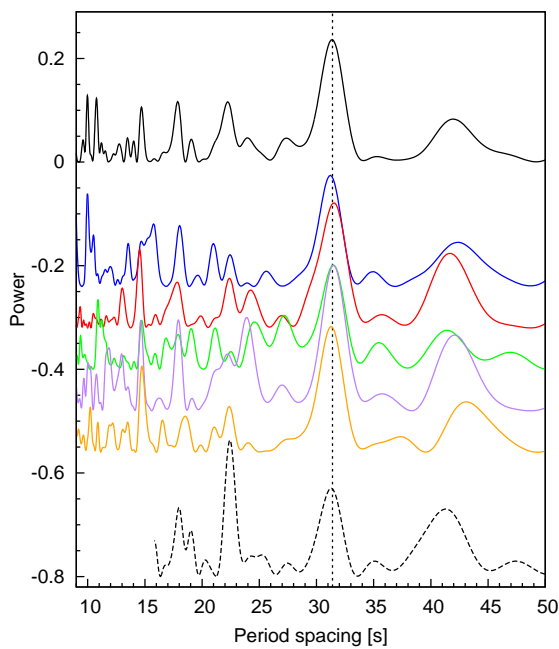}
\caption{Fourier transforms of different period sets. From top to bottom: FT of the periods of 
Table~\ref{table:finalfrek} (black solid line), test FTs of 5 of the 20 `aliased' period lists (coloured lines), 
FT of the combined list (black dashed line) using the periods determined by \citet{2003MNRAS.340.1031H} and five 
additional periods from the Konkoly2007 dataset (altogether 14 items). The vertical dotted line denotes the 31.4\,s 
spacing value.} 
\label{fig:tests2}
\end{figure}

Finally, we created a period list containing the periods presented in \citet{2003MNRAS.340.1031H} and marked in 
Table~\ref{table:finalfrek}, which were selected as the dominant and well-separated peaks from the different epochs' 
datasets. We complemented this with five periods from the Konkoly2007 dataset (686.6, 615.9, 
594.6, 439.1 and 395.1\,s), and also performed their Fourier analysis. This list contains 14 periods only, but we still obtain a 
maximum at 31.3\,s in the $\Delta\Pi=25-40$\,s region. Figure~\ref{fig:tests2} shows the FT of the original period 
set of Table~\ref{table:finalfrek}, five of the FTs of the aliased period lists, and the FT of the 14-period list.

These tests demonstrate that even if the dataset suffers from alias problems, the results from different 
methods remain consistent and thus the mean period spacing at 31\,s is a robust finding. Considering all of the mean period 
spacings derived by the linear fits, the K-S, I-V and Monte Carlo (Fourier) tests, we consider $31.4 \pm 0.3$\,s 
as the best estimation for the mean spacing of $\ell=1$ modes. 

\section{Stellar models}

\subsection{New spectroscopic mass of \mbox{KUV 05134+2605}}

We employed the DB white dwarf evolutionary tracks of \citet{2012A&A...541A..42C} to infer the stellar mass of 
\mbox{KUV 05134+2605} from the spectroscopic determination of $T_\mathrm{eff} = 24\,700\pm1300$\,K and $\log g = 8.21 \pm 0.06$ 
by \citet{2011ApJ...737...28B}.
These tracks cover a wide range of effective temperatures ($35\,000 \gtrsim T_{\rm eff} \gtrsim 18\,000$\,K) and 
masses ($0.51 \lesssim M_*/M_{\odot} \lesssim 0.870$). The sequences of DB white dwarf models have 
been obtained with a complete treatment of the evolutionary history of progenitors stars (see \citealt{2009ApJ...704.1605A}
for details or \citealt{2012A&A...541A..42C} for a brief discussion).
We varied the stellar mass and the effective temperature parameters in our model calculations, while 
the helium content, the chemical structure at the core, and the thickness of the chemical transition regions were 
fixed by the evolutionary history of progenitor objects.

\begin{figure}
\centering
\includegraphics[width=8.3cm]{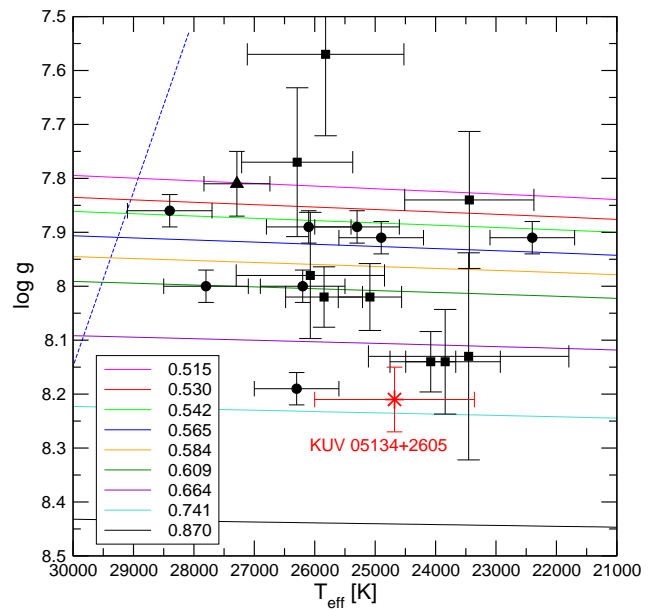}
\caption{The location of the known DBV stars on the $T_{\rm eff} - \log g$ 
plane. Also included are our DB white dwarf evolutionary tracks displayed
with different colours according to the stellar mass. The location of
\mbox{KUV 05134+2605} as given by spectroscopy is emphasized with a red asterisk
symbol. The theoretical blue edge of the DBV instability strip corresponding to the version MLT2
($\alpha = 1.25$) of the MLT theory of convection, as derived by \citet{2009JPhCS.172a2075C}, is 
depicted with a blue dashed line.} 
\label{tracks}
\end{figure}

In Fig.~\ref{tracks} we plot 
the evolutionary tracks along with the location of all the DBV stars known to date. 
We infer a new value of the spectroscopic mass for this star on the basis of this set of evolutionary models. 
This is relevant because these same set of DB white dwarf models will be utilized in the next sections to derive 
the stellar mass from the period spacing and the individual pulsation periods of \mbox{KUV 05134+2605}. 
We found $M_*= 0.72\pm0.04\,M_{\odot}$, in agreement with the value quoted in Sect.~\ref{sect:pp} ($M_*= 0.73\pm0.04\,M_{\odot}$).

\subsection{Clues of the stellar mass from the period spacing}

The observed mean period spacing of the dipole modes of \mbox{KUV 05134+2605},
as derived and checked in Sect.~\ref{sect:testspp}, is 
$\Delta \Pi_{\ell= 1}= 31.4 \pm 0.3$\,s. 
This period spacing is certainly shorter than the values measured in other DBV stars, 
that is, EC20058-5234 ($\Delta \Pi= 36.5$\,s; \citealt{2011ApJ...742L..16B}), 
GD 358 ($\Delta \Pi= 38.8$\,s; \citealt{2009ApJ...693..564P}), 
and WD J1929+4447 ($\Delta \Pi= 35.9$\,s; \citealt{2011ApJ...742L..16B}). 
This suggests that \mbox{KUV 05134+2605} could be more massive than other pulsating stars 
within the V777 Her class.

\begin{figure} 
\centering 
\includegraphics[width=8.3cm]{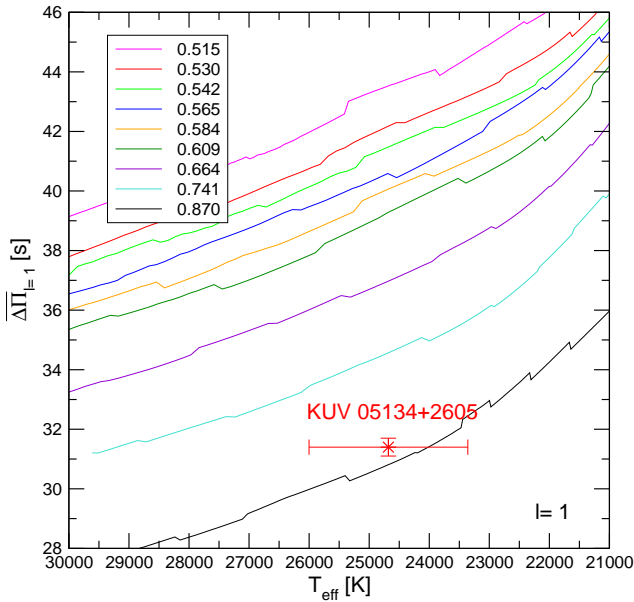} 
\caption{The average of the computed period spacings for $\ell= 1$ 
corresponding to our DB white dwarf sequences with different stellar 
masses. The location of \mbox{KUV 05134+2605} is shown with a red asterisk 
symbol ($T_{\rm eff} = 24\,700\pm1300$\,K) for the observed 
$\Delta \Pi_{\ell= 1}= 31.4 \pm 0.3$ solution of mean period spacing. 
By simple linear interpolation, we found that the mass of the star according 
to its period spacing is $M_*= 0.85 \pm 0.05\,M_{\odot}$.} 
\label{apsp-l1} 
\end{figure}

In Fig.~\ref{apsp-l1}, we show the 
run of the average of the computed period
spacings [Eq.~(\ref{avgdp})] with $\ell= 1$, 
in terms of the effective temperature for all of
our DB white-dwarf evolutionary sequences. The $g$-mode pulsation periods
were computed with the \textsc{LP-PUL} adiabatic pulsation code described in
\citet{2006A&A...454..863C}. The curves displayed in
the plot are somewhat jagged. This is because the average of the
computed period spacings is evaluated for a fixed period interval, and
not for a fixed $k$-interval. As the star evolves towards lower
effective temperatures, the periods generally increase with time. At
a given $T_{\rm eff}$, there are $n$ computed periods laying in the
chosen period interval. Later, when the model has cooled enough, it
is possible that the accumulated period drift nearly matches the
period separation between adjacent modes ($|\Delta k|= 1$). In these
circumstances, the number of periods laying in the chosen (fixed)
period interval is $n \pm 1$, and $\overline{\Delta \Pi}$
exhibits a small jump. In order to smooth the curves of
$\overline{\Delta \Pi}$, in constructing Fig.~\ref{apsp-l1}, we
have considered pulsation periods in a wider range of periods
($100\lesssim \Pi_k \lesssim 1200$\,s) than that observed in 
\mbox{KUV 05134+2605} ($390 \lesssim \Pi_k \lesssim 780$\,s)\footnote{If we 
adopt a shorter range of periods, closer to the range of periods exhibited 
by \mbox{KUV 05134+2605} (say $390-780$\,s), the curves we 
obtain are much more irregular and jumped, although the results 
do not appreciable change.}. The location of \mbox{KUV 05134+2605} is 
indicated by a red asterisk symbol according to the obtained solution for 
$\Delta \Pi_{\ell=1}$ (Sect.~\ref{sect:testspp}). 

We can have an estimate of the stellar mass free from any possible uncertainty 
in the spectroscopic analysis.  If we look for the possible range of masses for the star 
to be within the DBV instability strip ($21\,500 \lesssim T_{\rm eff} \lesssim 29\,000$ K), 
we find that $0.74 \lesssim M_*/M_{\odot} \lesssim 1.00$. This conclusion relies only on the measured 
mean period spacing, and on the average of the computed
period spacing, a global quantity that is not affected by the precise
shape of the chemical transition regions of the models. It is
particularly worth noting that this result is independent of the mass
of the He-rich envelope. Indeed, \citet{1990ApJS...72..335T} have shown that
the asymptotic period spacing of DBV white dwarfs is particularly
insensitive to the mass of the He-rich mantle, changing less than $1$\,s
for a change of 10 orders of magnitude in the mass of the He-rich
mantle (see their Fig.~42). Also, this conclusion is independent of
the effective temperature of the star. So, the result that
the mass of \mbox{KUV 05134+2605} is higher than the average mass of 
DB white dwarfs (see \citealt{2007MNRAS.375.1315K, 2011ApJ...737...28B, 2013ApJS..204....5K})
is a robust finding.

If we adopt the
spectroscopic effective temperature of \mbox{KUV 05134+2605} 
($T_{\rm eff} = 24\,700\pm1300$\,K) as a constraint, we infer a mass for the star 
of $M_*= 0.85 \pm 0.05\,M_{\odot}$. This value is larger than the spectroscopic mass 
($\sim 0.72-0.73\,M_{\odot}$). Note that both estimates of the stellar mass of 
\mbox{KUV 05134+2605} are based on the same DB white dwarf evolutionary models, 
which is a necessary condition in order that the comparison between the two values to be consistent.

\subsection{Constraints from the individual observed periods}
\label{sect:models}

The usual way to infer the stellar mass, the 
effective temperature and also details of the internal structure 
of pulsating white dwarfs is investigating their individual pulsation periods. 
In this approach, we seek a pulsation DB white dwarf model 
that best matches the pulsation periods of \mbox{KUV 05134+2605}. The 
goodness of the match between the theoretical pulsation periods ($\Pi_k$) 
and the observed individual periods ($\Pi_{{\rm obs}, i}$) is measured by 
means of a quality function defined as: 
\begin{equation}
\chi^2(M_*, T_{\rm eff})= \frac{1}{N} \sum_{i=1}^{N} 
\min[(\Pi_{{\rm obs},i}- \Pi_k)^2], 
\label{ptpf}
\end{equation}
\noindent where $N$ is the number of observed periods. The DB
white dwarf model that shows the lowest value of $\chi^2$ is adopted
as the `best-fit model'. Our asteroseismological approach combines
(i) a significant exploration of the parameter space, and (ii) a detailed
and updated input physics, in particular, regarding the internal structure, 
that is a crucial aspect for correctly disentangling the information encoded
in the pulsation patterns of variable white dwarfs (see 
\citealt{2007A&A...461.1095C, 2007A&A...475..619C, 2008A&A...478..869C, 
2009A&A...499..257C, 2012A&A...541A..42C, 2012MNRAS.420.1462R, 2013ApJ...779...58R}). 

We evaluate the function
$\chi^2(M_*, T_{\rm eff})$ for stellar masses of $0.515, 0.530, 0.542,
0.565, 0.584, 0.609, 0.664, 0.741$, and $0.870\,M_{\odot}$. For the
effective temperature, we employed a finer grid ($\Delta
T_{\rm eff}= 10-30$\,K). The quality of our period fits is assessed by
means of the average of the absolute period differences,
$\overline{\delta}= (\sum_{i=1}^N |\delta_i|)/N$, where $\delta_i=
\Pi_{{\rm obs}, i} -\Pi_k$, and by the root-mean-square residual,
$\sigma= \sqrt{(\sum |\delta_i|^2)/N}= \sqrt{\chi^2}$.

Given the unusually large (for DBV standards) number of pulsation
periods observed in \mbox{KUV 05134+2605}, 
we have to consider several cases regarding the set of periods to be 
employed in our asteroseismological fits.

\subsubsection{Case~0}

Here, we consider that all the identified periods shown 
in Table~\ref{table:finalfrek} are associated with either 
$\ell= 1$ \emph{or} $\ell= 2$, i.e., no mixture of $\ell$.
Given the presence of very small separations between 
some periods in the pulsation spectrum of \mbox{KUV 05134+2605},
then the modes should be all $\ell= 2$. However, it is improbable that
a pulsating white dwarf shows only quadrupole modes \citep{1994ApJ...430..839W, 2008A&A...477..627C},
and we discard it.

\subsubsection{Case~1}

In this case, we assume that the periods exhibited by 
\mbox{KUV 05134+2605} correspond to a mix of dipole and quadrupole
modes. Here, we perform a period fit in which 
the value of $\ell$ for the theoretical periods
is not fixed at the outset but instead is obtained as an 
output of our period fit procedure, with the allowed values 
of $\ell= 1$ and $\ell= 2$. We assume as real the existence of five rotational 
multiplets of frequencies and we select the period of the 
$m= 0$ component (if present) to perform our period 
fit, and the period corresponding to the average of the frequencies
supposed to be the $m= -1$ and $m= +1$ components 
if the $m= 0$ component is absent. We also assume that
the component $m= 0$ in the supposed triplet at periods
in the interval $525.8 - 528.5$\,s is the 
period at 525.8\,s\footnote{We have also considered the case
in which the period adopted is 528.5\,s, and also the situation
in which the pair $525.8, 528.5$ is not really a multiplet, but 
instead they are two distinct modes with different $\ell$-values. We have 
obtained almost identical results than those described in this Section,
just like \citet{2003BaltA..12..247M}.}. 
Specifically, we adopt a set of 16 periods.

We investigated the quantity $(\chi^2)^{-1}$ in terms of the effective temperature for 
different stellar masses, and found one strong maximum for a model 
with $M_*= 0.87\,M_{\odot}$ and $T_{\rm eff}= 25\,980$\,K. 
Such a pronounced maximum in the inverse of $\chi^2$ implies a 
good agreement between the theoretical and observed 
periods. The effective temperature of this model is somewhat
higher than (but still in agreement with) the spectroscopic effective
temperature of \mbox{KUV 05134+2605} ($T_{\rm eff} =
24\,700\pm1300$\,K). Another maximum, albeit much less pronounced, is
encountered for a model with the same stellar mass and lower
effective temperature ($T_{\rm eff} \sim 24\,000$\,K). However,
because the agreement between the observed and theoretical periods for
these models are much poorer, we adopt the model with $T_{\rm eff}=
25\,980$\,K as the best-fit asteroseismological model. 
In order to have an indicator of the quality of the period fit, we computed the
Bayes Information Criterion (BIC; \citealt{2000MNRAS.311..636K}):
\begin{equation}
{\rm BIC}= N_{\rm p} \left(\frac{\log N}{N} \right) + \log \sigma^2,
\end{equation}
\noindent where $N_{\rm p}$ is the number of free parameters, and $N$
is the number of observed periods. The smaller the value of BIC, the
better the quality of the fit. In our case, $N_{\rm p}= 2$ (stellar
mass and effective temperature), $N= 16$, and $\sigma= 2.8$\,s. 
We obtained ${\rm BIC}= 1.06$, which 
means that our fit is relatively good.

An important caveat should be kept in mind: from the period-fit
procedure, most of the periods are associated with $\ell= 2$ modes. This
is in contradiction with the well-known property that $\ell= 1$
modes exhibit substantially larger amplitudes than $\ell= 2$ ones,
because geometric cancellation effects become increasingly severe as
$\ell$ increases \citep{1977AcA....27....1D}. As a result of this,
pulsating white dwarfs should exhibit preferentially dipole modes.
Finally, we note a shortcoming of our asteroseismological model that
consists in that the pair of close periods at (395.1, 399.2)\,s is
fitted by a single period at 398.1\,s.

\subsubsection{Case~2}

Here, we assume that the periods close to each other (the `pairs' of
periods) that cannot be considered as components of triplets with a
frequency separation of $\sim 9\,\mu$Hz are actually a single period,
and then we consider the period resulting from the average of the
frequencies present. That is, we replace the pair (395.1, 399.2)\,s
by a single period at 397.2\,s, the pair (679.1, 686.6)\,s by a period
at 682.8\,s, and the pair (706.8, 719.1)\,s by a period at 712.9\,s.
The `reduced' list contains 13 periods. In this case, the best-fit
model has $M_*= 0.87\,M_{\odot}$ and $T_{\rm eff}= 25\,890$\,K
$[(\chi^2)^{-1}\sim0.13]$, thus the solution found in {\it Case~1}
still persists. However, in this case, other solutions [$T_{\rm eff}
 \sim 27\,100$ K, $(\chi^2)^{-1}\sim0.12$, and $T_{\rm eff} \sim
 24\,150$ K, $(\chi^2)^{-1}\sim0.11$] with the same stellar mass are
equally valid from a statistical point of view. The number of $\ell =
2$ modes still exceeds the number of $\ell = 1$ modes in the fit.

\subsubsection{Case~3}

\begin{figure} 
\centering 
\includegraphics[viewport=0 0 705 480, clip=true, width=8.6cm]{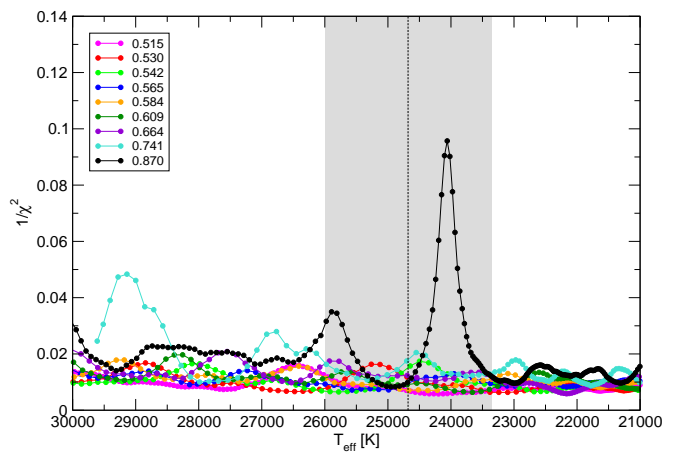} 
\caption{The inverse of the quality function of the period fit in
 terms of the effective temperature for {\it Case~3}. The
 vertical grey strip indicates the spectroscopic $T_{\rm eff}$ and
 its uncertainties ($T_{\rm eff} = 24\,700\pm1300$\,K).}
\label{fig:case3} 
\end{figure}

At variance with the previous cases, here we assume the
$\ell$-identification shown in Table~\ref{table:finalfrek} to be
correct and then we constrain the $\ell = 1$ observed modes to be
fitted by $\ell = 1$ theoretical modes, and the $\ell = 2$ observed
modes to be fitted by $\ell = 2$ theoretical modes. Note that,
according to Table~\ref{table:finalfrek}, there are several possible
assumptions for the mode identification. Specifically, the period at
439.1\,s can be associated to $\ell= 1$ or $\ell= 2$; the $m= 0$
component in the (525.8, 528.5)\,s doublet  
can be either 525.8\,s or 528.5\,s; and the period at 679.1\,s may be
associated to $\ell= 1$ and the period at 686.6\,s to $\ell= 2$ or
vice versa. On the other hand, the period at 395.1\,s must be an $\ell=
2$ mode if the period at 399.2\,s is an $\ell= 1$ mode. Similarly, the
706.8\,s period must be associated to $\ell= 2$ if the 719.1\,s period
is and $\ell= 1$ mode. We have taken into account all the possible
combinations in our period fits. No matter what assumptions we adopt
for the periods at 439.1\,s, 525.8\,s, 528.5\,s, 679.1\,s, and 686.6\,s, we
obtain a clear and recurrent single solution for a model with $M_*=
0.87\,M_{\odot}$ and $T_{\rm eff}= 24\,060$\,K. The set of observed
periods and their assigned $\ell$ values (9 modes with $\ell= 1$ and
7 modes with $\ell= 2$) are listed in the first two columns of
Table~\ref{table_case_3}. The theoretical periods of this fit are
shown in the third column of this Table, and the results of the
period fit are displayed in Fig.~\ref{fig:case3}. The quality of the
fit is described by a BIC index of 1.17.

\begin{table}
\centering
\caption{Comparison between the observed 16 periods and theoretical
 ($\ell= 1, 2$) periods corresponding to the seismological solutions
 in {\it Case~3} (3rd and 4th columns), with $M_*= 0.87\,M_{\odot}$
 and $T_{\rm eff}= 24\,060$\,K, and considering the additional
 evolutionary sequences with masses between $M_*= 0.74\,M_{\odot}$
 and $M_*= 0.87\,M_{\odot}$ (5th and 6th columns, $\Pi_{k, zoom}$
 and $k_{zoom}$). In the latter case, the model has $M_*=
 0.84\,M_{\odot}$ and $T_{\rm eff}= 25\,050$\,K. Observed periods in
 italics are `fictitious' (average) periods (see text). The $\ell$
 values in the model fits are fixed according to the solutions shown
 in Table~\ref{table:finalfrek} and the lower panels of
 Fig.~\ref{fig:pfits}.}
\begin{tabular}{cccccc}
\hline\hline
$\Pi_{{\rm obs},i} $ & $\ell$ & $\Pi_{k, Case~3}$ & $k_{C3}$ & $\Pi_{k, zoom}$ & $k_{zoom}$\\
\hline
395.1       & 2 & 399.9 & 19 & 401.3 & 19  \\        
399.2       & 1 & 398.8 & 10 & 400.7 & 10  \\        
439.1       & 2 & 439.2 & 21 & 440.7 & 21  \\       
525.8       & 1 & 522.6 & 14 & 524.7 & 14  \\        
546.1       & 2 & 549.6 & 27 & 548.2 & 27  \\
{\it 558.6} & 1 & 555.2 & 15 & 557.2 & 15  \\
567.2       & 2 & 567.7 & 28 & 568.7 & 28  \\ 
{\it 591.4} & 1 & 590.3 & 16 & 591.1 & 16  \\
615.9       & 1 & 620.2 & 17 & 618.0 & 17  \\ 
{\it 649.1} & 1 & 644.0 & 18 & 649.9 & 18  \\  
664.7       & 2 & 668.2 & 33 & 666.4 & 33  \\
679.1       & 1 & 682.7 & 19 & 682.1 & 19  \\
686.6       & 2 & 689.5 & 34 & 685.6 & 34  \\
706.8       & 2 & 703.6 & 35 & 699.7 & 35  \\
719.1       & 1 & 717.4 & 20 & 716.9 & 20  \\
776.6       & 1 & 772.4 & 22 & 774.8 & 22  \\
\hline
\end{tabular}
\label{table_case_3}
\end{table}

\subsection{Zooming in on the best-fit model region}

To search for possible closer solutions in 
the region of the parameter space where the best-fit model from 
period-to-period fits ($M_* \sim 0.87\,M_{\odot}$) was found, we have 
extended our exploration by generating a set of additional DB white 
dwarf evolutionary sequences with masses between $0.74\,M_{\odot}$ and 
$0.87\,M_{\odot}$. These sequences were constructed by scaling the 
stellar mass from the $0.87\,M_{\odot}$ sequence at very high 
effective temperatures. It is worth noting that the internal chemical 
profiles (of utmost importance regarding the pulsation properties of 
white dwarfs) of these new sequences are consistently assessed through 
linear interpolation between the chemical profiles of the sequences 
with $0.74\,M_{\odot}$ and $0.87\,M_{\odot}$. 

\begin{figure} 
\centering \includegraphics[viewport=0 0 705 480, clip=true, width=8.6cm]{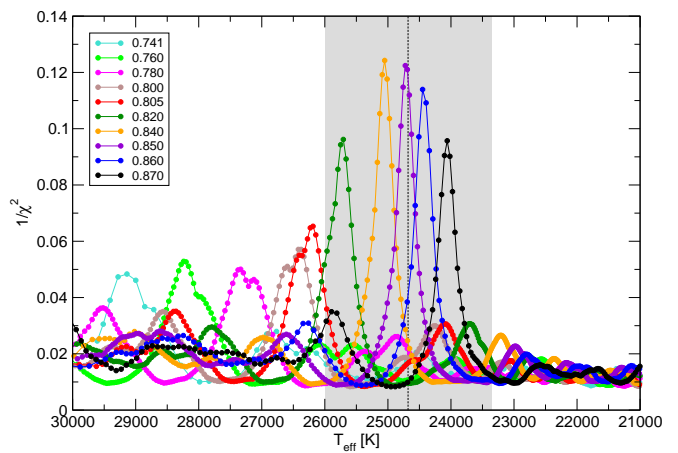}
\caption{Same as in Fig.~\ref{fig:case3}, but restricted to the set of model 
sequences with masses between $0.74\,M_{\odot}$ and $0.87\,M_{\odot}$.}
\label{case_3_zoom} 
\end{figure}

Specifically, we have computed $\ell= 1, 2$ $g$-mode periods of these
additional sequences with $M_*= 0.76, 0.78, 0.80, 0.805, 0.82, 0.84,
0.85, 0.86\,M_{\odot}$, and we have carried out period fits to these
models. We will focus the discussion on {\it Case~3}. 
The main results of the period fits are
displayed in Fig.~\ref{case_3_zoom}. The best solution found is a
model with a stellar mass of $0.84\,M_{\odot}$ and an effective
temperature of $25\,050$ K, characterized by a BIC index of 1.06, a
bit lower than for {\it Case~3}.

This period fit corresponds to the case in which the period at 439.1\,s 
is associated to $\ell= 1$, the $m= 0$ component in the (525.8, 528.5)\,s 
doublet is assumed to be at 525.8\,s, the
period at 679.1\,s is an $\ell= 1$, and the period at
686.6\,s is an $\ell= 2$ mode. We also searched for
asteroseismological solutions taking other combinations of these
assumptions into account. Interestingly, the assignation of the
harmonic degree to the periods at 679.1\,s and 686.6\,s proved to be
crucial for our results. Specifically, we obtain the solution
$(M_*/M_{\odot}, T_{\rm eff})= (0.84, 25\,050\,\mathrm{K})$ whenever we suppose
that the period at 679.1\,s is an $\ell= 1$, and the 686.6\,s period
is an $\ell= 2$ mode. If not, we obtain a second seismological
solution with $M_* = 0.82\,M_{\odot}$ and $T_{\rm eff}= 25\,700$\,K. 
The quality of the period fit, however, is worse (BIC= 1.19). 
We also performed fits replacing the pair (679.1, 686.6)\,s by a single period
at 682.8\,s. Assuming that this (average) period is $\ell= 1$, the best-fit 
solution is $M_* = 0.85\,M_{\odot}$ and $T_{\rm eff}= 24\,690$\,K. 
The preferred model is $M_* = 0.84\,M_{\odot}$ and $T_{\rm eff}= 25\,050$\,K, 
if the 682.8\,s period is $\ell= 2$.
Finally, we adopted the model with $(M_*/M_{\odot}, T_{\rm eff})= (0.84, 25\,050\,\mathrm{K})$, 
when the 679.1\,s period is an $\ell= 1$ and the 686.6\,s period
is an $\ell= 2$, as our asteroseismological solution for \mbox{KUV 05134+2605}.
This solution is characterized by the lowest BIC index, and has an effective temperature 
that is in excellent agreement with the 
value predicted by spectroscopy. The set of theoretical periods of this fit is listed
in Table~\ref{table_case_3} (fifth column). 

\subsubsection{The best-fit model}

The main features of our best-fit model considering the additional
evolutionary sequences are summarized in Table~\ref{table-best},
where we also include the parameters from spectroscopy
\citep{2011ApJ...737...28B}. In the Table, the quantity $M_{\rm He}$
corresponds to the total content of He of the envelope of the model.

\begin{figure}
\centering
\includegraphics[width=8.3cm]{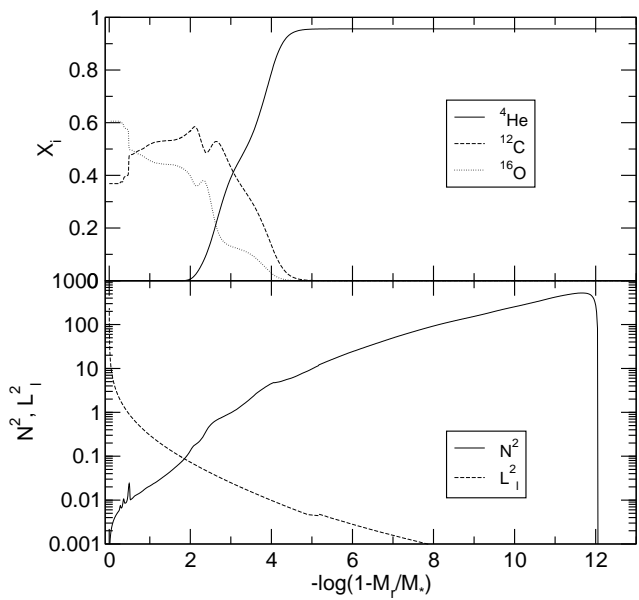}
\caption{The internal chemical structure (upper panel), and the
 squared Brunt-Va\"is\"al\"a and Lamb frequencies for $\ell= 1$
 (lower panel) corresponding to our best-fit DB white dwarf model
 with a stellar mass $M_*= 0.84\,M_{\odot}$, an effective temperature
 $T_{\rm eff}= 25\,050$\,K, and a He envelope mass of $M_{\rm{He}}/M_* \sim 1.5 \times 10^{-3}$.}
\label{x-b-n2}
\end{figure}

We briefly describe the main properties of our best-fit DB model for
\mbox{KUV 05134+2605}. In Fig.~\ref{x-b-n2}, we depict the internal chemical
structure of such model (upper panel), where the abundance by mass of
the main constituents ($^4$He, $^{12}$C, and $^{16}$O) is shown in
terms of the outer mass fraction [$-\log(1-M_r/M_*)$]. The chemical
structure of our models consists of a C/O core -- resulting from the
core He burning of the previous evolution -- shaped by processes of
extra mixing like overshooting. The core is surrounded by a mantle
rich in He, C, and O which is the remnant of the regions altered by
the nucleosynthesis during the thermally pulsing asymptotic giant
branch. Above this shell, there is a pure He mantle with a mass
$M_{\rm He}/M_* \sim 1.4 \times 10^{-3}$, constructed by the action of
gravitational settling that causes He to float to the surface and
heavier species to sink.

The lower panel of Fig.~\ref{x-b-n2} displays the run of the two
critical frequencies of non-radial stellar pulsations, that is, the
Brunt-V\"ais\"al\"a frequency and the Lamb frequency ($L_{\ell}$) for
$\ell= 1$. The precise shape of the Brunt-V\"ais\"al\"a frequency
largely determines the properties of the $g$-mode period spectrum of
the model. In particular, each chemical gradient in the model
contributes locally to the value of $N$. The most notable feature is
the very peaked structure at the C/O chemical transition
[$-\log(1-M_r/M_*) \sim 0.5$]. On the other hand, the He/C/O
transition region at $-\log(1-M_r/M_*) \sim 2-5$ is very smooth and
does not affect the pulsation spectrum much.

\begin{table}
\centering
\caption{The main characteristics of \mbox{KUV 05134+2605} and our best-fit 
model considering the additional evolutionary sequences with masses between 
$M_*= 0.74\,M_{\odot}$ and $M_*= 0.87\,M_{\odot}$. 
The second column corresponds to spectroscopic results, whereas the third
column present results from the best-fit asteroseismological model.}
\begin{tabular}{lrr}
\hline\hline
\noalign{\smallskip}
Quantity                  & Spectroscopy               & This work  \\
\hline
\noalign{\smallskip}
$T_{\rm eff}$ (K)            & $24\,700\pm1300^a$  &  $25\,050$  \\
$M_*$ ($M_{\odot}$)          & $0.72\pm 0.04$     &  $0.84$    \\ 
$\log g$ (cm/s$^2$)         & $8.21 \pm 0.06^a$     &  $8.37$    \\ 
$\log (L_*/L_{\odot})$       & \multicolumn{1}{c}{---} &  $-1.48$    \\  
$\log (R_*/R_{\odot})$        & \multicolumn{1}{c}{---} &  $-2.01$    \\  
$M_{\rm He}$ ($M_{\odot}$)    & \multicolumn{1}{c}{---} &  $1.18 \times 10^{-3}$ \\  
$X_{\rm O}$ (centre)     & \multicolumn{1}{c}{---} & $0.606$   \\
\noalign{\smallskip}
\hline
\multicolumn{3}{l}{$^{(a)}$\citet{2011ApJ...737...28B}}
\end{tabular}
\label{table-best}
\end{table}

\subsubsection{Periods from a DBA model}

\citet{2011ApJ...737...28B} reported that \mbox{KUV 05134+2605}
belongs to the DBA spectral class. Therefore, we performed
exploratory computations of the period spectrum of a DB white dwarf
model with pure helium atmosphere (`DB model') and a similar model
that has small impurities of hydrogen (`DBA model'). In the DBA
model, the hydrogen is mixed with helium all across the helium
envelope (like in the DBA models of \citealt{2009JPhCS.172a2075C},
sect.~4). Both models have $M_*= 0.74\,M_{\odot}$ and $T_{\rm eff}
\sim 24\,600$\,K. In particular, we have analysed the case in which
$\log(N_{\rm H}/N_{\rm He}) \sim -3.8$ (the value corresponding to
\mbox{KUV 05134+2605} according to \citealt{2011ApJ...737...28B}).

We have found that the relative differences in the pulsation periods
of these two models are in average lower than $0.2\%$. This means an
absolute difference lower than 0.8\,s for periods around 400\,s, and
below 1.6\,s for periods around 800\,s. That is, the presence of small
abundances of hydrogen in the almost pure helium atmospheres of DB
white dwarfs does not significantly alter the pulsation periods. As a
result, pure helium envelope models like the ones we employed in this
study are suitable for asteroseismic modelling of DBA white dwarfs
like \mbox{KUV 05134+2605}.

\section{Summary and conclusions}

The almost continuous, high-precision light curves provided by photometric space missions completely 
revolutionized the variable star studies in many fields. However, up to know only one 
V777 Her and three ZZ Ceti stars were monitored from space \citep{2014MNRAS.438.3086G, 2011ApJ...741L..16H, 
2011ApJ...736L..39O, 2014ApJ...789...85H}. This means that exploring their pulsational properties and thus studying their interiors 
are still mainly based on ground-based observations. Even thought their pulsation periods are short, in many 
cases extended observations (e.g. multisite campaigns and/or at least one season-long monitoring) are needed 
to determine enough (at least half dozen) normal modes for asteroseismic investigations (see e.g. the cases 
of KUV\,02464+3239, \citealt{2009MNRAS.399.1954B}, and \object{GD\,154}, \citealt{2013MNRAS.432..598P}). Otherwise, 
the low number of detected modes can result in models with fairly different physical parameters that can fit the 
observed periods similarly well. This shows the particular importance of the rich pulsators, in which at least 
a dozen normal modes can be fitted. Amplitude variations on short time scales seem to be common among the white 
dwarf pulsators. It can help to detect new modes, as observing the star in different epochs, we can find new 
frequencies and complement the list of known excited modes.

In the case of \mbox{KUV 05134+2605}, we used photometric datasets obtained in seven different epochs between 1988 
and 2011. We re-analysed the data published by \citet{2003MNRAS.340.1031H}, and also collected data at the 
Piszk\'estet\H{o} mountain station of Konkoly Observatory in two different observing seasons. Comparing the frequency 
content of the different datasets and considering the results of the frequency determination tests, 
we arrived at 22 pulsation frequencies between $1280$ and $2530\,\mu$Hz. 
The grouping of frequencies suggests at least 12 possible modes for the asteroseismic model fits. With this, 
\mbox{KUV 05134+2605} joined the -- not populous -- group of rich white dwarf pulsators. We also showed that considering 
a $\approx9\,\mu$Hz frequency separation, one triplet and at least one doublet can be determined. Assuming that 
the frequencies showing the $\approx9\,\mu$Hz 
separation are $\ell=1$ modes, we estimated the stellar rotation period as $0.6$\,d. We determined a 31\,s mean period 
spacing using different methods, as linear fits to the observed periods, Kolmogorov-Smirnov and Inverse Variance tests, 
and Fourier analysis of different period lists. We also simulated the effect of alias ambiguities on the mean period spacing 
determination. All of the results were consistent and point to the 31\,s characteristic spacing for the dipole modes. 

The extended list of frequencies provided the opportunity for asteroseismic investigations of \mbox{KUV 05134+2605} 
for the first time. In the case of period-to-period fits, the effective temperature of the selected model is 
$T_{\rm eff}= 25\,050$\,K, which is the same as the spectroscopic one ($T_{\rm eff}= 24\,700\pm1300$\,K, 
\citealt{2011ApJ...737...28B}) within the uncertainties. This model's helium layer mass is $1.2 \times 10^{-3}\,M_{\odot}$.

We found an excellent agreement between the stellar mass derived from the period spacing data and the 
period-to-period fits, all providing $M_*= 0.84-0.85\,M_{\odot}$ solutions. Note that the mass inferred from spectroscopy 
is $17-18\%$ lower than these values. One should keep in mind, however, that around the V777 Her instability strip, the 
helium lines are at maximum strength and therefore there is an ambiguity about a hot and a cool solution, just like 
the ambiguity of the high and low mass from the asteroseismological side. Also, there are still uncertainties in the 
synthetic line profile calculations. Similarly, we make assumptions on the physics of stars in the asteroseismic 
modelling. The discussion of which method is better to estimate the stellar mass and why, is out of the scope of this
paper.

Therefore, we consider the agreement between the results of three methods to be satisfactory. All three methods 
result a mass for \mbox{KUV 05134+2605} higher than the average mass of DB white dwarfs. Moreover, our study 
suggests that this star is the most massive one amongst the known DBVs.

\begin{acknowledgements}

The authors thank the anonymous referee for the constructive comments and recommendations on the manuscript.
We thank Gerald Handler for providing the \mbox{KUV 05134+2605} data obtained 
in 1988, 1992, 2000 October, 2001, and during the Whole Earth Telescope campaign in 2000 (XCov20). 
The authors also acknowledge the contribution of L. Moln\'ar, E. Plachy, H. Oll\'e and 
E. Vereb\'elyi to the observations of \mbox{KUV 05134+2605}. We thank \'Ad\'am S\'odor for critically 
reading the manuscript and his useful comments. Zs.B. acknowledges the support of the Hungarian E\"otv\"os 
Fellowship (2013) and the kind hospitality of the Royal Observatory of Belgium as a temporary voluntary researcher 
(2013--2014). Zs.~B. and M.~P. acknowledge the support of the ESA PECS project 4000103541/11/NL/KML.

\end{acknowledgements}




\begin{thebibliography}{61}
\expandafter\ifx\csname natexlab\endcsname\relax\def\natexlab#1{#1}\fi

\bibitem[{{Althaus} {et~al.}(2010){Althaus}, {C{\'o}rsico}, {Isern}, \&
  {Garc{\'{\i}}a-Berro}}]{2010A&ARv..18..471A}
{Althaus}, L.~G., {C{\'o}rsico}, A.~H., {Isern}, J., \& {Garc{\'{\i}}a-Berro},
  E. 2010, \aapr, 18, 471

\bibitem[{{Althaus} {et~al.}(2009){Althaus}, {Panei}, {Miller Bertolami},
  {Garc{\'{\i}}a-Berro}, {C{\'o}rsico}, {Romero}, {Kepler}, \&
  {Rohrmann}}]{2009ApJ...704.1605A}
{Althaus}, L.~G., {Panei}, J.~A., {Miller Bertolami}, M.~M., {et~al.} 2009,
  \apj, 704, 1605

\bibitem[{{Bergeron} {et~al.}(2011){Bergeron}, {Wesemael}, {Dufour},
  {Beauchamp}, {Hunter}, {Saffer}, {Gianninas}, {Ruiz}, {Limoges}, {Dufour},
  {Fontaine}, \& {Liebert}}]{2011ApJ...737...28B}
{Bergeron}, P., {Wesemael}, F., {Dufour}, P., {et~al.} 2011, \apj, 737, 28

\bibitem[{{Bischoff-Kim}(2009)}]{2009AIPC.1170..621B}
{Bischoff-Kim}, A. 2009, in American Institute of Physics Conference Series,
  Vol. 1170, American Institute of Physics Conference Series, ed. J.~A. {Guzik}
  \& P.~A. {Bradley}, 621--624

\bibitem[{{Bischoff-Kim} \& {Metcalfe}(2012)}]{2012ASPC..462..164B}
{Bischoff-Kim}, A. \& {Metcalfe}, T.~S. 2012, in Astronomical Society of the
  Pacific Conference Series, Vol. 462, Progress in Solar/Stellar Physics with
  Helio- and Asteroseismology, ed. H.~{Shibahashi}, M.~{Takata}, \& A.~E.
  {Lynas-Gray}, 164

\bibitem[{{Bischoff-Kim} \& {{\O}stensen}(2011)}]{2011ApJ...742L..16B}
{Bischoff-Kim}, A. \& {{\O}stensen}, R.~H. 2011, \apjl, 742, L16

\bibitem[{{Bogn{\'a}r} {et~al.}(2009){Bogn{\'a}r}, {Papar{\'o}}, {Bradley}, \&
  {Bischoff-Kim}}]{2009MNRAS.399.1954B}
{Bogn{\'a}r}, Zs., {Papar{\'o}}, M., {Bradley}, P.~A., \& {Bischoff-Kim}, A.
  2009, \mnras, 399, 1954

\bibitem[{{Bradley} \& {Winget}(1991)}]{1991ApJS...75..463B}
{Bradley}, P.~A. \& {Winget}, D.~E. 1991, \apjs, 75, 463

\bibitem[{{Bradley} {et~al.}(1993){Bradley}, {Winget}, \&
  {Wood}}]{1993ApJ...406..661B}
{Bradley}, P.~A., {Winget}, D.~E., \& {Wood}, M.~A. 1993, \apj, 406, 661

\bibitem[{{Castanheira} \& {Kepler}(2008)}]{2008MNRAS.385..430C}
{Castanheira}, B.~G. \& {Kepler}, S.~O. 2008, \mnras, 385, 430

\bibitem[{{C{\'o}rsico} \& {Althaus}(2006)}]{2006A&A...454..863C}
{C{\'o}rsico}, A.~H. \& {Althaus}, L.~G. 2006, \aap, 454, 863

\bibitem[{{C{\'o}rsico} {et~al.}(2008){C{\'o}rsico}, {Althaus}, {Kepler},
  {Costa}, \& {Miller Bertolami}}]{2008A&A...478..869C}
{C{\'o}rsico}, A.~H., {Althaus}, L.~G., {Kepler}, S.~O., {Costa}, J.~E.~S., \&
  {Miller Bertolami}, M.~M. 2008, \aap, 478, 869

\bibitem[{{C{\'o}rsico} {et~al.}(2012){C{\'o}rsico}, {Althaus}, {Miller
  Bertolami}, \& {Bischoff-Kim}}]{2012A&A...541A..42C}
{C{\'o}rsico}, A.~H., {Althaus}, L.~G., {Miller Bertolami}, M.~M., \&
  {Bischoff-Kim}, A. 2012, \aap, 541, A42

\bibitem[{{C{\'o}rsico} {et~al.}(2009{\natexlab{a}}){C{\'o}rsico}, {Althaus},
  {Miller Bertolami}, \& {Garc{\'{\i}}a-Berro}}]{2009A&A...499..257C}
{C{\'o}rsico}, A.~H., {Althaus}, L.~G., {Miller Bertolami}, M.~M., \&
  {Garc{\'{\i}}a-Berro}, E. 2009{\natexlab{a}}, \aap, 499, 257

\bibitem[{{C{\'o}rsico} {et~al.}(2009{\natexlab{b}}){C{\'o}rsico}, {Althaus},
  {Miller Bertolami}, \& {Garc{\'{\i}}a-Berro}}]{2009JPhCS.172a2075C}
{C{\'o}rsico}, A.~H., {Althaus}, L.~G., {Miller Bertolami}, M.~M., \&
  {Garc{\'{\i}}a-Berro}, E. 2009{\natexlab{b}}, Journal of Physics Conference
  Series, 172, 012075

\bibitem[{{C{\'o}rsico} {et~al.}(2007{\natexlab{a}}){C{\'o}rsico}, {Althaus},
  {Miller Bertolami}, \& {Werner}}]{2007A&A...461.1095C}
{C{\'o}rsico}, A.~H., {Althaus}, L.~G., {Miller Bertolami}, M.~M., \& {Werner},
  K. 2007{\natexlab{a}}, \aap, 461, 1095

\bibitem[{{C{\'o}rsico} {et~al.}(2007{\natexlab{b}}){C{\'o}rsico}, {Miller
  Bertolami}, {Althaus}, {Vauclair}, \& {Werner}}]{2007A&A...475..619C}
{C{\'o}rsico}, A.~H., {Miller Bertolami}, M.~M., {Althaus}, L.~G., {Vauclair},
  G., \& {Werner}, K. 2007{\natexlab{b}}, \aap, 475, 619

\bibitem[{{Costa} {et~al.}(2008){Costa}, {Kepler}, {Winget}, {O'Brien},
  {Kawaler}, {Costa}, {Giovannini}, {Kanaan}, {Mukadam}, {Mullally}, {Nitta},
  {Proven{\c c}al}, {Shipman}, {Wood}, {Ahrens}, {Grauer}, {Kilic}, {Bradley},
  {Sekiguchi}, {Crowe}, {Jiang}, {Sullivan}, {Sullivan}, {Rosen}, {Clemens},
  {Janulis}, {O'Donoghue}, {Ogloza}, {Baran}, {Silvotti}, {Marinoni},
  {Vauclair}, {Dolez}, {Chevreton}, {Dreizler}, {Schuh}, {Deetjen}, {Nagel},
  {Solheim}, {Gonzalez Perez}, {Ulla}, {Barstow}, {Burleigh}, {Good},
  {Metcalfe}, {Kim}, {Lee}, {Sergeev}, {Akan}, {{\c C}ak{\i}rl{\i}}, {Paparo},
  {Viraghalmy}, {Ashoka}, {Handler}, {H{\"u}rkal}, {Johannessen}, {Kleinman},
  {Kalytis}, {Krzesinski}, {Klumpe}, {Larrison}, {Lawrence}, {Mei{\v s}tas},
  {Martinez}, {Nather}, {Fu}, {Pak{\v s}tien{\.e}}, {Rosen},
  {Romero-Colmenero}, {Riddle}, {Seetha}, {Silvestri}, {Vu{\v c}kovi{\'c}},
  {Warner}, {Zola}, {Althaus}, {C{\'o}rsico}, \&
  {Montgomery}}]{2008A&A...477..627C}
{Costa}, J.~E.~S., {Kepler}, S.~O., {Winget}, D.~E., {et~al.} 2008, \aap, 477,
  627

\bibitem[{{Csubry} \& {Koll{\'a}th}(2004)}]{2004ESASP.559..396C}
{Csubry}, Z. \& {Koll{\'a}th}, Z. 2004, in ESA Special Publication, Vol. 559,
  SOHO 14 Helio- and Asteroseismology: Towards a Golden Future, ed.
  D.~{Danesy}, 396

\bibitem[{{Dziembowski}(1977)}]{1977AcA....27....1D}
{Dziembowski}, W. 1977, \actaa, 27, 1

\bibitem[{{Eastman} {et~al.}(2010){Eastman}, {Siverd}, \&
  {Gaudi}}]{2010PASP..122..935E}
{Eastman}, J., {Siverd}, R., \& {Gaudi}, B.~S. 2010, \pasp, 122, 935

\bibitem[{{Fontaine} \& {Brassard}(2008)}]{2008PASP..120.1043F}
{Fontaine}, G. \& {Brassard}, P. 2008, \pasp, 120, 1043

\bibitem[{{Fu} {et~al.}(2013){Fu}, {Dolez}, {Vauclair}, {Fox-Machado}, {Kim},
  {Li}, {Chen}, {Alvarez}, {Su}, {Charpinet}, {Chevreton}, {Michel}, {Yang},
  {Li}, {Zhang}, {Molnar}, \& {Plachy}}]{2013MNRAS.429.1585F}
{Fu}, J.-N., {Dolez}, N., {Vauclair}, G., {et~al.} 2013, \mnras, 429, 1585

\bibitem[{{Grauer} {et~al.}(1989){Grauer}, {Wegner}, {Green}, \&
  {Liebert}}]{1989AJ.....98.2221G}
{Grauer}, A.~D., {Wegner}, G., {Green}, R.~F., \& {Liebert}, J. 1989, \aj, 98,
  2221

\bibitem[{{Greiss} {et~al.}(2014){Greiss}, {G{\"a}nsicke}, {Hermes}, {Steeghs},
  {Koester}, {Ramsay}, {Barclay}, \& {Townsley}}]{2014MNRAS.438.3086G}
{Greiss}, S., {G{\"a}nsicke}, B.~T., {Hermes}, J.~J., {et~al.} 2014, \mnras,
  438, 3086

\bibitem[{{Handler} {et~al.}(2002){Handler}, {Metcalfe}, \&
  {Wood}}]{2002MNRAS.335..698H}
{Handler}, G., {Metcalfe}, T.~S., \& {Wood}, M.~A. 2002, \mnras, 335, 698

\bibitem[{{Handler} {et~al.}(2003){Handler}, {O'Donoghue}, {M{\"u}ller},
  {Solheim}, {Gonzalez-Perez}, {Johannessen}, {Paparo}, {Szeidl}, {Viraghalmy},
  {Silvotti}, {Vauclair}, {Dolez}, {Pallier}, {Chevreton}, {Kurtz}, {Bromage},
  {Cunha}, {{\O}stensen}, {Fraga}, {Kanaan}, {Amorim}, {Giovannini}, {Kepler},
  {da Costa}, {Anderson}, {Wood}, {Silvestri}, {Klumpe}, {Carlton}, {Miller},
  {McFarland}, {Grauer}, {Kawaler}, {Riddle}, {Reed}, {Nather}, {Winget},
  {Hill}, {Metcalfe}, {Mukadam}, {Kilic}, {Watson}, {Kleinman}, {Nitta},
  {Guzik}, {Bradley}, {Sekiguchi}, {Sullivan}, {Sullivan}, {Shobbrook},
  {Jiang}, {Birch}, {Ashoka}, {Seetha}, {Girish}, {Joshi}, {Dorokhova},
  {Dorokhov}, {Akan}, {Mei{\v s}tas}, {Janulis}, {Kalytis}, {Ali{\v s}auskas},
  {Anguma}, {Kalebwe}, {Moskalik}, {Ogloza}, {Stachowski}, {Pajdosz}, \&
  {Zola}}]{2003MNRAS.340.1031H}
{Handler}, G., {O'Donoghue}, D., {M{\"u}ller}, M., {et~al.} 2003, \mnras, 340,
  1031

\bibitem[{{Handler} {et~al.}(1997){Handler}, {Pikall}, {O'Donoghue}, {Buckley},
  {Vauclair}, {Chevreton}, {Giovannini}, {Kepler}, {Goode}, {Provencal},
  {Wood}, {Clemens}, {O'Brien}, {Nather}, {Winget}, {Kleinman}, {Kanaan},
  {Watson}, {Nitta}, {Montgomery}, {Klumpe}, {Bradley}, {Sullivan}, {Wu},
  {Marar}, {Seetha}, {Ashoka}, {Mahra}, {Bhat}, {Babu}, {Leibowitz}, {Hemar},
  {Ibbetson}, {Mashal}, {Meistas}, {Dziembowski}, {Pamyatnykh}, {Moskalik},
  {Zola}, {Pajdosz}, {Krzesinski}, {Solheim}, {Bard}, {Massacand}, {Breger},
  {Gelbmann}, {Paunzen}, \& {North}}]{1997MNRAS.286..303H}
{Handler}, G., {Pikall}, H., {O'Donoghue}, D., {et~al.} 1997, \mnras, 286, 303

\bibitem[{{Hermes} {et~al.}(2014){Hermes}, {Charpinet}, {Barclay}, {Pak{\v
  s}tien{\.e}}, {Mullally}, {Kawaler}, {Bloemen}, {Castanheira}, {Winget},
  {Montgomery}, {Van Grootel}, {Huber}, {Still}, {Howell}, {Caldwell}, {Haas},
  \& {Bryson}}]{2014ApJ...789...85H}
{Hermes}, J.~J., {Charpinet}, S., {Barclay}, T., {et~al.} 2014, \apj, 789, 85

\bibitem[{{Hermes} {et~al.}(2013){Hermes}, {Montgomery}, {Gianninas}, {Winget},
  {Brown}, {Harrold}, {Bell}, {Kenyon}, {Kilic}, \&
  {Castanheira}}]{2013MNRAS.436.3573H}
{Hermes}, J.~J., {Montgomery}, M.~H., {Gianninas}, A., {et~al.} 2013, \mnras,
  436, 3573

\bibitem[{{Hermes} {et~al.}(2011){Hermes}, {Mullally}, {{\O}stensen},
  {Williams}, {Telting}, {Southworth}, {Bloemen}, {Howell}, {Everett}, \&
  {Winget}}]{2011ApJ...741L..16H}
{Hermes}, J.~J., {Mullally}, F., {{\O}stensen}, R.~H., {et~al.} 2011, \apjl,
  741, L16

\bibitem[{{Jones} {et~al.}(1989){Jones}, {Hansen}, {Pesnell}, \&
  {Kawaler}}]{1989ApJ...336..403J}
{Jones}, P.~W., {Hansen}, C.~J., {Pesnell}, W.~D., \& {Kawaler}, S.~D. 1989,
  \apj, 336, 403

\bibitem[{{Kawaler}(1988)}]{1988IAUS..123..329K}
{Kawaler}, S.~D. 1988, in IAU Symposium, Vol. 123, Advances in Helio- and
  Asteroseismology, ed. J.~{Christensen-Dalsgaard} \& S.~{Frandsen}, 329

\bibitem[{{Kawaler} {et~al.}(1999){Kawaler}, {Sekii}, \&
  {Gough}}]{1999ApJ...516..349K}
{Kawaler}, S.~D., {Sekii}, T., \& {Gough}, D. 1999, \apj, 516, 349

\bibitem[{{Kepler} {et~al.}(2007){Kepler}, {Kleinman}, {Nitta}, {Koester},
  {Castanheira}, {Giovannini}, {Costa}, \& {Althaus}}]{2007MNRAS.375.1315K}
{Kepler}, S.~O., {Kleinman}, S.~J., {Nitta}, A., {et~al.} 2007, \mnras, 375,
  1315

\bibitem[{{Kepler} {et~al.}(2003){Kepler}, {Nather}, {Winget}, {Nitta},
  {Kleinman}, {Metcalfe}, {Sekiguchi}, {Jiang}, {Sullivan}, {Sullivan},
  {Janulis}, {Meistas}, {Kalytis}, {Krzesinski}, {Ogoza}, {Zola}, {O'Donoghue},
  {Romero-Colmenero}, {Martinez}, {Dreizler}, {Deetjen}, {Nagel}, {Schuh},
  {Vauclair}, {Ning}, {Chevreton}, {Solheim}, {Gonzalez Perez}, {Johannessen},
  {Kanaan}, {Costa}, {Murillo Costa}, {Wood}, {Silvestri}, {Ahrens}, {Jones},
  {Collins}, {Boyer}, {Shaw}, {Mukadam}, {Klumpe}, {Larrison}, {Kawaler},
  {Riddle}, {Ulla}, \& {Bradley}}]{2003A&A...401..639K}
{Kepler}, S.~O., {Nather}, R.~E., {Winget}, D.~E., {et~al.} 2003, \aap, 401,
  639

\bibitem[{{Kleinman} {et~al.}(2013){Kleinman}, {Kepler}, {Koester}, {Pelisoli},
  {Pe{\c c}anha}, {Nitta}, {Costa}, {Krzesinski}, {Dufour}, {Lachapelle},
  {Bergeron}, {Yip}, {Harris}, {Eisenstein}, {Althaus}, \&
  {C{\'o}rsico}}]{2013ApJS..204....5K}
{Kleinman}, S.~J., {Kepler}, S.~O., {Koester}, D., {et~al.} 2013, \apjs, 204, 5

\bibitem[{{Kleinman} {et~al.}(1998){Kleinman}, {Nather}, {Winget}, {Clemens},
  {Bradley}, {Kanaan}, {Provencal}, {Claver}, {Watson}, {Yanagida}, {Nitta},
  {Dixson}, {Wood}, {Grauer}, {Hine}, {Fontaine}, {Liebert}, {Sullivan},
  {Wickramasinghe}, {Achilleos}, {Marar}, {Seetha}, {Ashoka}, {Meistas},
  {Leibowitz}, {Moskalik}, {Krzesinski}, {Solheim}, {Bruvold}, {O'Donoghue},
  {Kurtz}, {Warner}, {Martinez}, {Vauclair}, {Dolez}, {Chevreton}, {Barstow},
  {Kepler}, {Giovannini}, {Augusteijn}, {Hansen}, \&
  {Kawaler}}]{1998ApJ...495..424K}
{Kleinman}, S.~J., {Nather}, R.~E., {Winget}, D.~E., {et~al.} 1998, \apj, 495,
  424

\bibitem[{{Koen} \& {Laney}(2000)}]{2000MNRAS.311..636K}
{Koen}, C. \& {Laney}, D. 2000, \mnras, 311, 636

\bibitem[{{Koll{\'a}th}(1990)}]{1990KOTN....1....1K}
{Koll{\'a}th}, Z. 1990, Konkoly Observatory Occasional Technical Notes, 1, 1

\bibitem[{{Lenz} \& {Breger}(2005)}]{2005CoAst.146...53L}
{Lenz}, P. \& {Breger}, M. 2005, Communications in Asteroseismology, 146, 53

\bibitem[{{Metcalfe}(2003)}]{2003BaltA..12..247M}
{Metcalfe}, T.~S. 2003, Baltic Astronomy, 12, 247

\bibitem[{{Montgomery} {et~al.}(2010){Montgomery}, {Provencal}, {Kanaan},
  {Mukadam}, {Thompson}, {Dalessio}, {Shipman}, {Winget}, {Kepler}, \&
  {Koester}}]{2010ApJ...716...84M}
{Montgomery}, M.~H., {Provencal}, J.~L., {Kanaan}, A., {et~al.} 2010, \apj,
  716, 84

\bibitem[{{Nather} {et~al.}(1990){Nather}, {Winget}, {Clemens}, {Hansen}, \&
  {Hine}}]{1990ApJ...361..309N}
{Nather}, R.~E., {Winget}, D.~E., {Clemens}, J.~C., {Hansen}, C.~J., \& {Hine},
  B.~P. 1990, \apj, 361, 309

\bibitem[{{O'Donoghue}(1994)}]{1994MNRAS.270..222O}
{O'Donoghue}, D. 1994, \mnras, 270, 222

\bibitem[{{{\O}stensen} {et~al.}(2011){{\O}stensen}, {Bloemen}, {Vu{\v
  c}kovi{\'c}}, {Aerts}, {Oreiro}, {Kinemuchi}, {Still}, \&
  {Koester}}]{2011ApJ...736L..39O}
{{\O}stensen}, R.~H., {Bloemen}, S., {Vu{\v c}kovi{\'c}}, M., {et~al.} 2011,
  \apjl, 736, L39

\bibitem[{{Papar{\'o}} {et~al.}(2013){Papar{\'o}}, {Bogn{\'a}r}, {Plachy},
  {Moln{\'a}r}, \& {Bradley}}]{2013MNRAS.432..598P}
{Papar{\'o}}, M., {Bogn{\'a}r}, Zs., {Plachy}, E., {Moln{\'a}r}, L., \&
  {Bradley}, P.~A. 2013, \mnras, 432, 598

\bibitem[{{Provencal} {et~al.}(2009){Provencal}, {Montgomery}, {Kanaan},
  {Shipman}, {Childers}, {Baran}, {Kepler}, {Reed}, {Zhou}, {Eggen}, {Watson},
  {Winget}, {Thompson}, {Riaz}, {Nitta}, {Kleinman}, {Crowe}, {Slivkoff},
  {Sherard}, {Purves}, {Binder}, {Knight}, {Kim}, {Chen}, {Yang}, {Lin}, {Lin},
  {Chen}, {Jiang}, {Sergeev}, {Mkrtichian}, {Andreev}, {Janulis}, {Siwak},
  {Zola}, {Koziel}, {Stachowski}, {Paparo}, {Bognar}, {Handler}, {Lorenz},
  {Steininger}, {Beck}, {Nagel}, {Kusterer}, {Hoffman}, {Reiff}, {Kowalski},
  {Vauclair}, {Charpinet}, {Chevreton}, {Solheim}, {Pakstiene}, {Fraga}, \&
  {Dalessio}}]{2009ApJ...693..564P}
{Provencal}, J.~L., {Montgomery}, M.~H., {Kanaan}, A., {et~al.} 2009, \apj,
  693, 564

\bibitem[{{Provencal} {et~al.}(2012){Provencal}, {Montgomery}, {Kanaan},
  {Thompson}, {Dalessio}, {Shipman}, {Childers}, {Clemens}, {Rosen},
  {Henrique}, {Bischoff-Kim}, {Strickland}, {Chandler}, {Walter}, {Watson},
  {Castanheira}, {Wang}, {Handler}, {Wood}, {Vennes}, {Nemeth}, {Kepler},
  {Reed}, {Nitta}, {Kleinman}, {Brown}, {Kim}, {Sullivan}, {Chen}, {Yang},
  {Shih}, {Jiang}, {Sergeev}, {Maksim}, {Janulis}, {Baliyan}, {Vats}, {Zola},
  {Baran}, {Winiarski}, {Ogloza}, {Paparo}, {Bognar}, {Papics}, {Kilkenny},
  {Sefako}, {Buckley}, {Loaring}, {Kniazev}, {Silvotti}, {Galleti}, {Nagel},
  {Vauclair}, {Dolez}, {Fremy}, {Perez}, {Almenara}, \&
  {Fraga}}]{2012ApJ...751...91P}
{Provencal}, J.~L., {Montgomery}, M.~H., {Kanaan}, A., {et~al.} 2012, \apj,
  751, 91

\bibitem[{{Reed} {et~al.}(2011){Reed}, {Baran}, {Quint}, {Kawaler}, {O'Toole},
  {Telting}, {Charpinet}, {Rodr{\'{\i}}guez-L{\'o}pez}, {{\O}stensen},
  {Provencal}, {Johnson}, {Thompson}, {Allen}, {Middour}, {Kjeldsen}, \&
  {Christensen-Dalsgaard}}]{2011MNRAS.414.2885R}
{Reed}, M.~D., {Baran}, A., {Quint}, A.~C., {et~al.} 2011, \mnras, 414, 2885

\bibitem[{{Reegen}(2007)}]{2007A&A...467.1353R}
{Reegen}, P. 2007, \aap, 467, 1353

\bibitem[{{Romero} {et~al.}(2012){Romero}, {C{\'o}rsico}, {Althaus}, {Kepler},
  {Castanheira}, \& {Miller Bertolami}}]{2012MNRAS.420.1462R}
{Romero}, A.~D., {C{\'o}rsico}, A.~H., {Althaus}, L.~G., {et~al.} 2012, \mnras,
  420, 1462

\bibitem[{{Romero} {et~al.}(2013){Romero}, {Kepler}, {C{\'o}rsico}, {Althaus},
  \& {Fraga}}]{2013ApJ...779...58R}
{Romero}, A.~D., {Kepler}, S.~O., {C{\'o}rsico}, A.~H., {Althaus}, L.~G., \&
  {Fraga}, L. 2013, \apj, 779, 58

\bibitem[{{Tassoul} {et~al.}(1990){Tassoul}, {Fontaine}, \&
  {Winget}}]{1990ApJS...72..335T}
{Tassoul}, M., {Fontaine}, G., \& {Winget}, D.~E. 1990, \apjs, 72, 335

\bibitem[{{Van Grootel} {et~al.}(2013){Van Grootel}, {Fontaine}, {Brassard}, \&
  {Dupret}}]{2013ApJ...762...57V}
{Van Grootel}, V., {Fontaine}, G., {Brassard}, P., \& {Dupret}, M.-A. 2013,
  \apj, 762, 57

\bibitem[{{Vauclair} {et~al.}(2011){Vauclair}, {Fu}, {Solheim}, {Kim}, {Dolez},
  {Chevreton}, {Chen}, {Wood}, {Silver}, {Bogn{\'a}r}, {Papar{\'o}}, \&
  {C{\'o}rsico}}]{2011A&A...528A...5V}
{Vauclair}, G., {Fu}, J.-N., {Solheim}, J.-E., {et~al.} 2011, \aap, 528, A5

\bibitem[{{Winget} \& {Kepler}(2008)}]{2008ARA&A..46..157W}
{Winget}, D.~E. \& {Kepler}, S.~O. 2008, \araa, 46, 157

\bibitem[{{Winget} {et~al.}(1991){Winget}, {Nather}, {Clemens}, {Provencal},
  {Kleinman}, {Bradley}, {Wood}, {Claver}, {Frueh}, {Grauer}, {Hine}, {Hansen},
  {Fontaine}, {Achilleos}, {Wickramasinghe}, {Marar}, {Seetha}, {Ashoka},
  {O'Donoghue}, {Warner}, {Kurtz}, {Buckley}, {Brickhill}, {Vauclair}, {Dolez},
  {Chevreton}, {Barstow}, {Solheim}, {Kanaan}, {Kepler}, {Henry}, \&
  {Kawaler}}]{1991ApJ...378..326W}
{Winget}, D.~E., {Nather}, R.~E., {Clemens}, J.~C., {et~al.} 1991, \apj, 378,
  326

\bibitem[{{Winget} {et~al.}(1994){Winget}, {Nather}, {Clemens}, {Provencal},
  {Kleinman}, {Bradley}, {Claver}, {Dixson}, {Montgomery}, {Hansen}, {Hine},
  {Birch}, {Candy}, {Marar}, {Seetha}, {Ashoka}, {Leibowitz}, {O'Donoghue},
  {Warner}, {Buckley}, {Tripe}, {Vauclair}, {Dolez}, {Chevreton}, {Serre},
  {Garrido}, {Kepler}, {Kanaan}, {Augusteijn}, {Wood}, {Bergeron}, \&
  {Grauer}}]{1994ApJ...430..839W}
{Winget}, D.~E., {Nather}, R.~E., {Clemens}, J.~C., {et~al.} 1994, \apj, 430,
  839

\bibitem[{{Winget} {et~al.}(1982){Winget}, {Robinson}, {Nather}, \&
  {Fontaine}}]{1982ApJ...262L..11W}
{Winget}, D.~E., {Robinson}, E.~L., {Nather}, R.~D., \& {Fontaine}, G. 1982,
  \apjl, 262, L11

\bibitem[{{Zima}(2008)}]{2008CoAst.155...17Z}
{Zima}, W. 2008, Communications in Asteroseismology, 155, 17

\end{thebibliography}


\listofobjects

\end{document}